\documentclass[12pt]{article}

\usepackage[margin=1in]{geometry}
\usepackage{setspace}
\usepackage{fancyhdr}
\usepackage{lastpage}
\usepackage{amsmath, amssymb, amsthm, bm, bbm}
\usepackage{graphicx}
\usepackage{hyperref}
\usepackage[inline]{enumitem}
\usepackage{xcolor}
\usepackage[labelfont={color=Rosso, bf},textfont=sc]{caption}
\usepackage{array}  
\usepackage{titlesec}  
\usepackage{tikz}  
\usepackage[english]{babel}
\usepackage[latin1]{inputenc}
\usepackage[T1]{fontenc}
\usepackage{graphicx}
\usepackage{multirow}
\usepackage{bm}
\usepackage[most]{tcolorbox}
\usepackage{array}
\usepackage{arydshln} 
\usepackage{wrapfig}  
\usepackage{float}    
\usepackage{needspace} 
\usepackage{natbib}

\hypersetup{
  colorlinks = true,
  linkcolor = Rosso,
  citecolor = Blu,
  urlcolor = blue
}

\def\balpha{{\bm\alpha}}

\def\emr{\pi_{pf}} \def\demr{\widetilde{\pi}_{pf}}
 
\def\py{p(y)}   \def\fya{f(y|\balpha)}

\def\scn{\mathcal{S}} \def\scnj{\scn_j} 

\def\safehaven{backstop } \def\safehavennospace{backstop}

\def\bem#1{{\em #1}}

\def\black{\color{black}}
\def\grigio{\color{lightgray}}
\def\rosso{\color{Rosso}}

\def\Y{\mathbf{Y}}

\def\y{\mathbf{y}}

\def\seq#1#2{#1{:}#2}
\def\j1J{j=\seq 1J}

\def\tN{\textrm{N}}
\def\tS{\textrm{S}}
\def\tT{\textrm{T}}

\def\skt{skew-\textit{t} }
\def\pdf{p.d.f. } 
\def\cdf{c.d.f. } 

\def\bmat{\begin{pmatrix}}
\def\emat{\end{pmatrix}}

\def\beq#1{\begin{equation}\label{eq:#1}} \def\eeq{\end{equation}}
\def\bea{\begin{align}}                   \def\eea{\end{align}}
\def\beas{\begin{align*}}                 \def\eeas{\end{align*}}

\DeclareMathOperator*{\argmax}{arg\,max}

\newcommand{\ben}{\begin{enumerate}} \newcommand{\een}{\end{enumerate}}
\newcommand{\bit}{\begin{itemize}}   \newcommand{\eit}{\end{itemize}}
\newcommand{\ba}{\begin{array}}      \newcommand{\ea}{\end{array}}

\newcommand{\mbf}{\mathbf}
\newcommand{\bbm}{\begin{bmatrix}} \newcommand{\ebm}{\end{bmatrix}}

\definecolor{Rosso}{RGB}{142,0,28}
\definecolor{Blu}{rgb}{0.0, 0.14, 0.4}

\titleformat{\section}
   {\color{Rosso}\titlerule[0.75pt]\normalfont\large\bfseries}
   {\thesection}{1em}{}
   [{\titlerule[0.75pt]}]

\titleformat{\subsection}
   {\color{Rosso}\normalfont\normalsize\bfseries}
   {\thesubsection}{1em}{}
   [{\titlerule[0.2pt]}]

\titleformat{\subsubsection}
   {\color{Rosso}\normalfont\normalsize\it}
   {\thesubsubsection}{1em}{}
   [{\titlerule[0.1pt]}]

\titleformat{\paragraph}[runin]
   {\color{Rosso}\normalfont\normalsize\sc}
   {\theparagraph}{1em}{}
   []
\renewenvironment{abstract}{%
  \begin{center}
   {\color{Rosso}\rule{\textwidth}{0.2pt}} \\[-3pt]
    {\color{Rosso}\normalfont\normalsize\bfseries Abstract}\\[-9pt]
    {\color{Rosso}\rule{\textwidth}{0.2pt}}
  \end{center}
   \vspace{-9pt}\small
}{%
  \par
}

\titlespacing*{\section}{0pt}{18pt}{9pt}
\titlespacing*{\subsection}{0pt}{12pt}{6pt}
\titlespacing*{\subsubsection}{0pt}{9pt}{6pt}
\titlespacing*{\paragraph}{0pt}{3pt}{1em}

\newcolumntype{L}[1]{>{\raggedright\let\newline\\\arraybackslash\hspace{0pt}}m{#1}}
\newcolumntype{C}[1]{>{\centering\let\newline\\\arraybackslash\hspace{0pt}}m{#1}}
\newcolumntype{R}[1]{>{\raggedleft\let\newline\\\arraybackslash\hspace{0pt}}m{#1}}

\graphicspath{{Figures/}}

\begin{document}

\renewcommand{\thefootnote}{$\ast$} 

\begin{center} 
\color{Rosso}
\rule{\textwidth}{1.5pt}\\
\textbf{\Huge Risks and Uncertainty\footnotemark\\[6pt] in Monetary Policy}\\
\rule{\textwidth}{1pt}\\ [18pt]  

\color{black}
\begin{tabular}{cp{2cm}c}
\normalsize \sc Tobias Adrian &&  \normalsize \sc Domenico Giannone \\[-5pt]
\small International Monetary Fund &&\small Johns Hopkins University\\[-5.5pt]
\footnotesize tadrian@imf.org &&\footnotesize domenico.giannone@jhu.edu\\[9pt]
\normalsize \sc Matteo Luciani &&  \normalsize \sc Mike West\\[-5pt]
\small Federal Reserve Board &&\small Duke University\\[-5.5pt]
\footnotesize matteo.luciani@frb.gov &&\footnotesize mike.west@duke.edu\\[9pt]
\end{tabular}

\today
\end{center}

\footnotetext{We are grateful for valuable comments and suggestions from Marta Ba\'{n}bura, Thomas Carter, Matyas Farkas, Manuel Gonzalez-Astudillo, Edward Herbst, David Hofman, Ruy Lama, Jesper Lind\'e, Simon Lloyd, Shalva Mkhatrishvili, Steve Mulema, Pau Rabanal, Yuji Sakurai, Giovanni Sciacovelli, and Dima Solohub.\\[2pt]
Disclaimer: The views expressed in this paper are those of the authors and do not necessarily reflect the views and policies of the Board of Governors, the Federal Reserve System, or the International Monetary Fund, its Management, or its Executive Directors.}

\renewcommand{\thefootnote}{\arabic{footnote}}

\begin{abstract}
Central banks monitor macroeconomic risk through two traditions: scenario analysis, regularly used since the mid-1990s, and distributional forecasting, practiced since the late 1960s. The two are complementary but separate: scenarios provide narratives without probabilities, while predictive distributions provide probabilities with limited economic interpretation. Treating baseline forecasts and scenarios as conditional predictive densities, and distributional forecasts as reference predictive distributions, places both within a common framework and clarifies their roles. The Scenario Synthesis assigns weights to scenarios consistent with the reference distribution, offering a practical and reproducible tool for risk assessment and policy deliberation under deep uncertainty.\\[-4pt]
\noindent \rosso{\rule{\textwidth}{0.2pt}} \\[6pt]
\noindent \rosso{\emph{Keywords}:} \black{Scenarios, fan charts, growth-at-risk, model uncertainty, Bayesian predictive synthesis.} \\
\rosso{\emph{JEL Codes}:} \black{C1, C11, C53, E32, E37, E58.}
\end{abstract}

\newpage

%
%

\section{Introduction}\label{sec:introduction}

\begin{figure}[ht]
\caption{The Two Approaches to Risk in the December 2018 Tealbook}\label{fig:TB2018}
\centering \smallskip
\begin{tabular}{C{.39\textwidth}C{.02\textwidth}C{.59\textwidth}}
\footnotesize \sc (a) Alternative Scenarios && \footnotesize \sc (b) Time-Varying Macroeconomic Risk \\
\scriptsize Real GDP Growth, 4-quarter percent change && \scriptsize GDP Growth forecast error, Percentage points\\[3pt]
\vspace{10pt}\includegraphics[width=.39\textwidth,trim=2.5cm 12cm 12cm 7.1cm, clip]{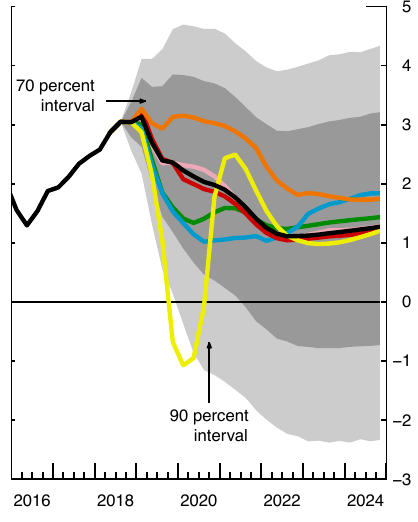} &&
\begin{tabular}{p{.59\textwidth}}
\includegraphics[width=.59\textwidth,trim=2.5cm 11.5cm 6.75cm 11.53cm, clip]{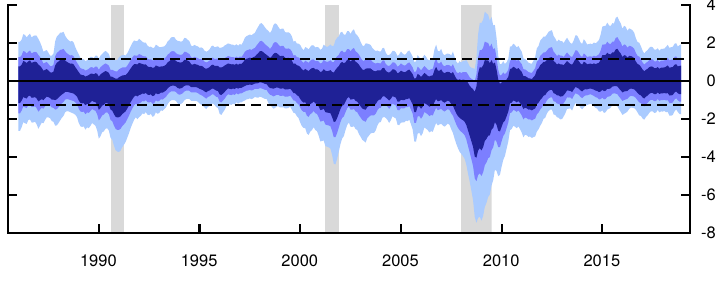} \\[-10pt]

\scriptsize
\textsc{Note:} Panel (\textsc{a}) reproduces the GDP panel from the Tealbook exhibit ``Forecast Confidence Intervals and Alternative Scenarios.'' The solid black line is the staff baseline projection, the shaded bands are 70 and 90 percent confidence intervals from FRB/US stochastic simulations, and the colored lines trace the alternative scenarios considered in December 2018, including a financial-based recession, stronger supply side, supply constraints, greater interest-rate sensitivity, foreign slowdown, and lower oil prices. Panel (\textsc{b}) reproduces the GDP panel from the exhibit ``Time-Varying Macroeconomic Risk.'' The shaded areas represent the predictive distribution of four-quarter-ahead Tealbook GDP growth forecast errors, conditional on indicators of real activity, inflation, financial market strain, and the volatility of high-frequency macroeconomic indicators; the straight (dashed) line marks the median (15th and 85th percentiles) of the unconditional distribution, while gray bars indicate NBER recessions. 
\end{tabular}
\\
\end{tabular}
\begin{tabular}{p{\textwidth}}
\scriptsize
\textsc{Source:} Both panels are reproduced from the ``Risks and Uncertainty'' chapter of the December 2018 Tealbook.
\end{tabular}
\end{figure}

Central banks have monitored macroeconomic risk systematically for roughly three decades. Staff regularly brief policy committees on the risks surrounding the economic outlook. At the Federal Reserve, this practice is evident in the Tealbook, the briefing document on economic and financial conditions that staff prepare for the Federal Open Market Committee (FOMC) ahead of each policy meeting. Since 2010, every Tealbook has contained a chapter titled ``Risks and Uncertainty'' that assesses the forces that could drive the economy away from the baseline projection. Figure~\ref{fig:TB2018} shows the two main tools Federal Reserve Board staff use to describe and communicate risk in the December 2018 Tealbook. Panel~(\textsc{a}) presents the \emph{alternative scenarios}, a few fully articulated paths for the economy built on internally consistent assumptions about shocks and their transmission mechanisms. Panel~(\textsc{b}) presents a \emph{predictive distribution} for GDP growth centered on the staff's baseline.

Despite appearing in the same chapter, the two exhibits in Figure~\ref{fig:TB2018} are not connected through a formal framework. The gap is conceptual, not merely organizational. The scenarios provide economic narratives but no probabilities: they describe how a financial-based recession might unfold, but not how likely it is. The predictive distribution provides probabilities but no economic interpretation: it shows that downside risk has risen sharply, but not why. Each tool answers the question the other leaves open. A policymaker who asks ``what is the probability of a financial-based recession?'' or ``what narrative underlies this tail risk?'' finds no answer in either exhibit.

This paper makes three contributions. First, it provides a historical account of how and why central banks developed the two approaches---tracing the intellectual origins of scenario thinking and probabilistic forecasting, the institutional pressures that shaped them, and why they developed in parallel rather than in dialogue. Second, it develops a formal framework that bridges them---the Scenario Synthesis---which assigns probabilities to named scenarios consistently with the predictive distribution and recovers economic narratives from probabilistic evidence, allowing the two exhibits to speak to each other. Third, it shows empirically that the framework gives central banks a disciplined way to characterize the balance of risks, assess whether scenario sets span the relevant sources of macroeconomic risk, and attach probabilities to named risks under clear conditions.

The historical survey reveals that these two approaches developed in parallel for distinct institutional and intellectual reasons, and that strong views persist about which one is superior. \citet{bernanke2024}, for example, in a review of the Bank of England's forecasting framework, criticized its fan charts---a format for communicating predictive distributions---as difficult to interpret and communicate, and recommended greater reliance on narrative scenarios. However, as \citet{Baueretal2025} observe in their review for the Fed's 2025 strategy assessment, ``no clear set of best practices for communicating uncertainty and risks to the public has emerged.'' We argue that this debate rests on a false dichotomy: scenarios and predictive densities are natural complements, and the challenge is to combine their strengths rather than to choose between them. This complementarity reflects a deeper reality: central banks face ``deep uncertainty''---about the model itself, not only about shocks---so the two tools approximate different objects and serve different purposes.

We formalize this distinction by treating baseline forecasts and scenarios as \emph{conditional} predictive densities: each is a forecast of the economy conditional on a specific set of assumptions and a specific model. The reference predictive distribution, instead, aims at approximating the \emph{unconditional} distribution---the probability of all outcomes, not conditioned on any particular model or set of assumptions. The Scenario Synthesis then asks whether this unconditional distribution can be approximated as a weighted combination of conditional views and, if so, what weights each scenario should receive.

We apply the Scenario Synthesis framework to Tealbook data from two contrasting episodes---the turbulent pre-crisis environment of December 2007 and the more stable conditions of December 2018. In 2007, the scenario set provided only limited coverage of downside risks to GDP growth (the Credit Crunch scenario captured meaningful probability mass, but the 2008--09 outcome had an even fatter left tail), and alternative reference densities implied sharply different weights. In 2018, by contrast, the scenario set was much closer to spanning the relevant risks and yielded an internally coherent assessment. More broadly, the exercises show how the Synthesis can be used to characterize the balance of risks, assess whether the available scenario set spans the relevant sources of macroeconomic risk, and, when it does not, guide the design of improved scenarios.

Beyond staff forecasting, the framework helps structure disagreement within policy committees. Under deep uncertainty, policymakers may disagree about shocks, transmission mechanisms, or the weighting of risks. By representing alternative views as conditional predictive densities and aggregating them into a committee-level assessment, the Scenario Synthesis makes these sources of disagreement explicit---and thus serves not only to assign probabilities to scenarios but to discipline policy deliberation.

The rest of the paper is organized as follows. Sections~\ref{sec:fed_history} and \ref{sec:other_cbs} trace the development of risk analysis at the Federal Reserve and other central banks. Section~\ref{sec:disconnect} discusses the disconnect between scenarios and predictive densities. Sections~\ref{sec:toolkit} and \ref{sec:scenario_synthesis} present the framework and the Scenario Synthesis. Section~\ref{sec:empirics} contains the empirical applications to the 2007 and 2018 Tealbooks. Sections~\ref{sec:learnings}, \ref{sec:economics}, and \ref{sec:policy} discuss the broader implications for scenario design, alternative economic models, and policy deliberation. Section~\ref{sec:conclusions} concludes.

%
%

\section{Risk Analysis at the Federal Reserve: A Historical Account}\label{sec:fed_history}

The systematic monitoring of macroeconomic risk by central banks dates back only to the 1990s. The practice became more important after the Global Financial Crisis and has gained further prominence since the COVID-19 pandemic. Yet documentation of these practices is scarce and fragmented. In the spirit of \citet{Sims2002BPEA}, this section reconstructs the Federal Reserve's risk-analysis practices and traces their evolution, drawing on publicly available material from the Tealbook, prepared by Federal Reserve Board (FRB) staff for the FOMC,\footnote{The Tealbook replaced the Greenbook, which served the same function from the 1960s through 2009. To simplify exposition, we use ``Tealbook'' throughout, while preserving ``Greenbook'' in archival titles, source notes, and citations.} and the Blackbook, prepared by the Federal Reserve Bank of New York for its President. Both documents are classified when prepared and released with a five-year lag.

\begin{figure}[ht]
\caption{Scenarios and Predictive Densities for December 2007}\label{fig:TB2007}
\centering \smallskip
\begin{tabular}{C{.39\textwidth}C{.02\textwidth}C{.59\textwidth}}
\footnotesize \sc (a) Alternative Scenarios && \footnotesize \sc (b) Probability Densities \\
\scriptsize Real GDP Growth, 4-quarter percent change && \scriptsize 2008/2007 Real GDP Growth Probabilities\\ [3pt]
\includegraphics[width=.39\textwidth,trim=2.5cm 12cm 12cm 6.75cm, clip]{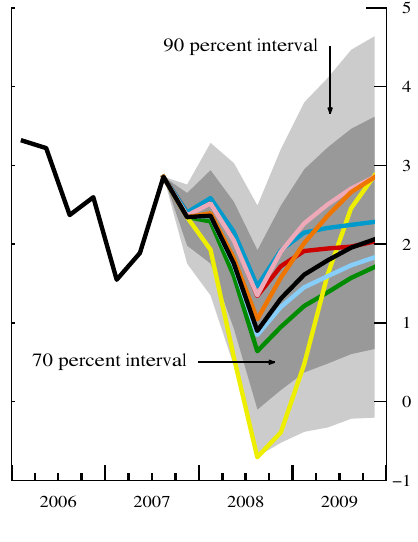}  &&
\vspace{-15pt}
\begin{tabular}{p{.59\textwidth}}
\includegraphics[width=.59\textwidth,trim=11cm 11cm 2cm 11.6cm, clip]{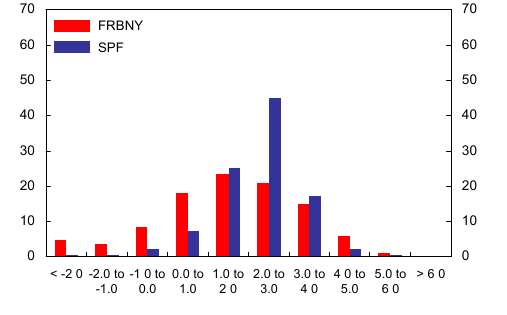} \\[-10pt]
\scriptsize
\textsc{Note:} Panel (\textsc{a}) shows the Tealbook baseline projection (thick black line) with seven alternative scenarios (colored lines) and confidence intervals from FRB/US stochastic simulations. Panel (\textsc{b}) shows histogram probabilities for 2008/2007 GDP growth:
red bars represent FRBNY staff judgmental assessment, blue bars represent SPF consensus forecast. 
\\[-2pt]
\scriptsize
\textsc{Source:} Panel~(\textsc{a}) is reproduced from the December 5, 2007 Greenbook. Panel~(\textsc{b}) is reproduced from the December 7, 2007 NY Fed Blackbook (p.~67). 
\end{tabular}
\\
\end{tabular}

\end{figure}

Two broad approaches have emerged. The first identifies salient economic risks and uses models to produce alternative scenarios---conditional forecasts describing what happens if those risks materialize. The second constructs predictive distributions over possible future outcomes using surveys, judgment, or statistical models. This distinction is central to the paper. The Introduction showed scenarios and predictive densities side by side for December 2018 (Figure~\ref{fig:TB2018}), a useful ``normal-times'' benchmark. Here we add an earlier, more revealing episode: December 2007, one year before the Global Financial Crisis. Figure~\ref{fig:TB2007} displays the Tealbook scenarios alongside the predictive densities from the NY Fed Blackbook and the Survey of Professional Forecasters. The contrast between these assessments lies at the heart of this paper. We begin with scenarios, turn to predictive densities, and then discuss the gap between them.

\subsection{The Baseline}\label{sec:fed_baseline}

The natural starting point is the Tealbook baseline, because both scenarios and predictive densities measure risk around it. The baseline is the FRB staff's judgmental projection under a particular set of conditioning assumptions, and it anchors the Fed's risk analysis. The Tealbook baseline is not the forecast of any single model. It combines sectoral expertise, incoming data, conditioning assumptions, econometric models, financial-market information, outside forecasts, and judgmental add-factors \citep{Fischer2017}.\footnote{Public Tealbooks explicitly contrast the staff forecast with model forecasts. In the December 2018 ``Alternative Model Forecasts'' exhibit, for example, the staff projection for 2019 real GDP growth is 2.4 percent versus 1.1 percent for FRB/US \citep[][p.~91]{Dec2018Tealbook}.} When, in Section~\ref{sec:toolkit}, we associate the baseline with a ``baseline model'' $M_0$, we mean the baseline process that anchors the staff projection, not the mechanical output of FRB/US.

Uncertainty around the baseline is summarized in several ways. The confidence intervals in Figures~\ref{fig:TB2018}(\textsc{a}) and \ref{fig:TB2007}(\textsc{a}) come from stochastic simulations of FRB/US, the Federal Reserve Board's workhorse semi-structural model \citep{BraytonTinsley1996,BraytonLevinTryonWilliams1997,brayton2014}. FRB/US, a large-scale model in the Cowles Foundation tradition, contains hundreds of equations covering consumption, investment, housing, labor markets, prices, fiscal and monetary policy, and international linkages, combining micro-founded and reduced-form elements; its main strengths are comprehensiveness and interpretability. Since its introduction in 1996, it has been used extensively to communicate ideas to the Board and the FOMC and to produce domestic alternative scenarios \citep{TetlowIronside2007}. The Tealbook also reports intervals derived from historical forecast errors, which provide an alternative way to summarize uncertainty around the baseline projection.\footnote{See the exhibits ``Prediction Intervals Derived from Historical Tealbook Forecast Errors'' (since 2017) and ``Selected Tealbook Projections and 70 Percent Confidence Intervals Derived from Historical Tealbook Forecast Errors and FRB/US Simulations'' (since 2004).}

Both approaches have important limitations. Because FRB/US is most often used in a (conditionally) linearized form, its simulated uncertainty via stochastic simulations is close to symmetric.\footnote{FRB/US is not itself a linear model, and even its linearized version respects the effective lower bound. The Board's stochastic-simulation procedure has been refined to generate more realistic recession depth, duration, and left-tail behavior by \citet{GonzalezAstudilloVilan2019}, who introduced a Markov-switching residual procedure. The resulting asymmetries are modest for GDP growth and inflation but more pronounced for the fed funds rate and unemployment.} Historical forecast-error intervals can be more asymmetric, but both approaches are time-invariant: they adjust only after shocks appear in the data, and cannot respond to contemporaneous indicators of risk such as current financial conditions \citep{Baueretal2025}. For this reason, the Fed's broader assessment of risk complements baseline intervals with alternative scenarios and time-varying predictive densities, discussed next.

\subsection{Monitoring Risk Using Scenarios}\label{sec:fed_scenarios}
A scenario is a forecast produced under alternative assumptions and, sometimes, an alternative model relative to the baseline. The model can be a formal econometric model or a ``mental model''---a qualitative narrative translated into a quantitative path through expert judgment. Each scenario is typically accompanied by a narrative explaining the economic mechanisms: which transmission channels are activated, why the assumptions lead to different outcomes, and how the scenario unfolds.

\paragraph{Scenarios at the Federal Reserve Board.}
At the Federal Reserve Board, combining baseline projections with alternative scenarios has been standard since the mid-1990s. Until 2009 these scenarios appeared in the U.S.\ outlook section of the Greenbook; since 2010 they have appeared in the Tealbook's ``Risks and Uncertainty'' section. The scenario set changes from round to round, depending on which risks staff consider most salient.

Early on, the Board constructed virtually all domestic scenarios using FRB/US \citep{HerbstKonzemScofield2026}. It later expanded the model set to include SIGMA \citep{SIGMA}, a multicountry DSGE model for international scenarios; EDO \citep{EDO}, an estimated DSGE model of the U.S.\ economy; and models with richer financial sectors, such as Gertler--Karadi-type credit-accelerator frameworks \citep{GertlerKaradi2011}.

\paragraph{Scenarios at the NY Fed.}
Scenarios have appeared in the Blackbook since at least 2005, historically built from expert judgment. By the December 2017 Blackbook---the last publicly available one---the process had become more model-based: scenarios were built using a medium-scale DSGE model \citep{DelNegroGiannoni2017} combined with a large BVAR \citep{crumpetal2025} estimated on the 31 variables in the Tealbook judged most important. The DSGE supplies a structural interpretation, while the BVAR, though reduced-form, is closely tied to the structural models and serves as a cross-check and a flexible tool for conditional forecasting.

Unlike the FRB staff, FRBNY staff also assigned judgmental weights to each scenario and combined them into a full probability distribution. This anticipates a key element of the Scenario Synthesis---the explicit weighting of scenarios---except that the Blackbook weights were judgmental, whereas the Synthesis chooses them statistically via a concordance criterion.

\paragraph{Limitations of scenarios.}
Scenarios provide narrative structure---economic mechanisms, transmission channels, and causal stories policymakers can reason about---but they offer no probability assessment and no formal metric to evaluate them as a set. Two questions therefore go unanswered: \emph{What is the probability of each scenario?} and \emph{How well does the scenario set capture current macroeconomic risks?} These are not merely conceptual concerns: when presented with an adverse scenario, policymakers naturally ask whether it warrants a response or a contingency plan, and the answer depends on information the scenario itself does not provide.

A second limitation is scenario selection. From round to round there is often little continuity: new scenarios appear, old ones disappear, and most persist only briefly. There is typically no formal criterion for which risks are explored, nor a systematic record of why a scenario is introduced or removed, making it hard to compare assessments over time or to tell whether changes reflect the economy or judgment. This critique is not unique to scenarios: statistical risk tools are themselves evolving research products rather than fixed reference objects.\footnote{For example, in April 2019 the staff introduced a new two-year macro-risk exhibit because the one-year ``Time-Varying Macroeconomic Risk'' exhibit could show balanced risks while staff perceived larger downside risks further out \citep[][pp.~75, 78]{Apr2019Tealbook}. During COVID-19, the staff suspended the two-year exhibit, judging its distributions unreliable.}

A third limitation is that scenarios are typically reported as point paths, with no predictive distribution; and even when models characterize uncertainty around them, those models are usually linear or log-linear, so that uncertainty within each scenario is typically symmetric and time-invariant, with little scope for fat tails, skewness, or state-dependent amplification. In practice, the risk assessment comes primarily from the choice of scenarios and the assumptions imposed on them, not the models' internal dynamic.

In sum, scenarios provide rich narratives but remain deeply judgmental---in which risks are explored, the assumptions defining them, and the models quantifying them. What is missing is a systematic way to evaluate whether the scenario set as a whole captures the relevant risks. The Scenario Synthesis of \citet{AGLW2025} fills this gap by leveraging information from predictive densities.

\subsection{Monitoring Risk Using Predictive Densities}\label{sec:fed_densities}

A predictive density assigns probabilities to all possible future realizations of a variable, describing the full range of outcomes---including particularly favorable or adverse ones---rather than delivering only a point forecast. A point forecast compresses the distribution into a single number that is often asked to carry two objects at once: what is most likely and what matters for decision-making. Econometricians define a point forecast as the minimizer of expected loss under a given loss function \citep{GrangerMachina2006,ElliottTimmermann2008,Gneiting2011}.\footnote{Expected loss under a predictive density is the statistical concept of risk. In this paper we use a broader notion of risk: the probabilistic assessment of possible future events.} Predictive densities separate the two: the density describes the risks, and the loss function weights them.

\paragraph{The Survey of Professional Forecasters.}
The Survey of Professional Forecasters (SPF) is the longest-running continuous source of predictive densities for the U.S. economy. Launched in 1968 by the American Statistical Association and the National Bureau of Economic Research, and administered by the Philadelphia Fed since 1990, it asks forecasters for point forecasts and for probabilities assigned to bins covering possible GDP growth realizations. Averaging these probabilities across respondents yields a consensus density summarizing the community's collective assessment. The blue bars in Figure~\ref{fig:TB2007}b show this consensus density for December 2007.

\paragraph{Predictive densities at the NY Fed.}
The New York Fed has produced predictive densities in its internal Blackbook since at least 2005, though the Blackbook's existence and methodology have been sparsely documented even within the Federal Reserve System. From May 2007 to March 2010, the New York Fed Research Group presented density forecasts using histograms in a format similar to the SPF, plotted side by side with SPF distributions. Figure~\ref{fig:TB2007}(\textsc{b}) shows an example from the December 2007 Blackbook: FRBNY staff assigned non-negligible probability mass to negative GDP growth outcomes, while the SPF remained centered on positive growth---illustrating how internal density forecasts can highlight downside tail risks that may be less visible in surveys. The next subsection discusses why the NY Fed's assessment differed so sharply from the SPF.

This exercise later became more formalized. In March 2017, the Blackbook featured SEPIA (Summary of Economic Projections with Individual Assessment of Uncertainty), an internal survey of about 13 Research Group economists. Each economist assigned probabilities to outcome bins, and averaging these probabilities produced a consensus density analogous to the SPF. SEPIA thus provided an internal, survey-based assessment of macroeconomic risks that complemented the Blackbook's other risk tools.

\paragraph{Why the NY Fed's risk assessment differed from the SPF.}
The contrast in Figure~\ref{fig:TB2007}(\textsc{b}) is striking. One year before the Great Recession, the NY Fed assigned roughly 15\% probability to negative GDP growth, while the SPF consensus remained near-symmetric with limited downside risk---a difference reflecting the NY Fed's distinct informational environment. 

The NY Fed houses the markets desk, which executes open-market operations and maintains daily contact with primary dealers, and it supervises the largest bank holding companies, giving it direct visibility into balance sheets, leverage, and funding conditions. This proximity to Wall Street gave it real-time intelligence about emerging stress in money markets, mortgage-backed securities, and interbank lending that was not captured by the macroeconomic data available to the SPF.

Remarks by NY Fed president Timothy Geithner at the August 7, 2007 FOMC meeting illustrate this advantage. He warned that the turmoil had ``the potential to cause substantial damage through the effects on asset prices, market liquidity, and credit; through the potential failure of more-consequential financial institutions; and through a general erosion of confidence,'' and noted that ``a lot of that risk has gone to leveraged funds that have much less capacity to absorb this kind of shock'' \citep[][p.~60]{FOMCTranscriptAug2007}. Cautioning more inflation-focused colleagues, he added that waiting for clearer evidence on aggregate demand would make them ``inevitably be too late'' \citep[][p.~62]{FOMCTranscriptAug2007}. The December 2007 Blackbook confirms this picture: FRBNY staff wrote that ``downside risks to growth have increased significantly,'' with credit spreads at post-2001 highs and interbank markets deteriorating beyond their August peaks \citep[p.~20]{Dec2007Blackbook}. The NBER later dated the recession to that month.

In short, the NY Fed used financial conditions---credit spreads, leverage, and funding markets---as a signal of downside risk that the SPF, relying primarily on traditional macroeconomic data, did not capture.

\paragraph{From judgment to econometrics: the Growth-at-Risk framework.}
A major development in the analysis of macroeconomic risk is the Growth-at-Risk (GaR) framework of \citet{Adrianetal2019}. Developed at the New York Fed to update forecast distributions at the roughly six- to seven-week FOMC frequency, it formalizes the link between financial conditions and macroeconomic risk through the entire conditional distribution of future GDP growth, not just the mean. Tighter financial conditions sharply worsen downside risk while leaving the upper part of the distribution comparatively stable: lower GDP-growth quantiles respond strongly to credit spreads, equity volatility, and leverage, upper quantiles much less. Financial conditions thus matter mainly for the tails.\footnote{The original framework \citep{Adrianetal2019}, which used financial conditions as the main driver, was later extended to economic uncertainty \citep{TBTVMR,KEIJSERS2025748} and geopolitical uncertainty \citep{Investment-at-Risk}.} This asymmetry is consistent with nonlinear financial-accelerator mechanisms, discussed in Section~\ref{sec:economics}, in which deteriorating balance sheets, widening risk premia, and credit contraction amplify downside risk. 

At the Federal Reserve Board, this line of work entered the Tealbook in 2017 through the Time-Varying Macroeconomic Risk (TVMR) exhibit (Figure~\ref{fig:TB2018}b). TVMR reports a predictive distribution for Tealbook forecast errors, with dispersion and asymmetry driven by indicators of real activity, inflation, economic uncertainty, and financial conditions \citep[see][]{TBTVMR}. In stress episodes, it implies wider downside risks; in more tranquil periods, such as late 2018, the distribution is closer to symmetric and centered around the baseline.

At the New York Fed, the same line of work is reflected in Outlook-at-Risk (OaR), published since 2023. OaR provides one-year-ahead predictive distributions for key macroeconomic variables and builds on extensions of the Growth-at-Risk framework by \citet{Adamsetal2021} and \citet{BoyarchenkoCrumpEliasLopezGaffney2026}, who construct risk distributions around consensus forecasts for GDP growth, unemployment, and inflation.\footnote{Related predictive-density methods are now an active area of research throughout the Federal Reserve System, with recent applications to unemployment, inflation, corporate spreads, tail risks in BVARs, financial shocks, and investment-at-risk \citep{Kiley2022,LopezSalidoLoria2024,CaldaraScottiZhong2021,CarrieroClarkMarcellino2024,Scotti2023,AmburgeyMcCracken2025}.}

Taken together, TVMR and OaR illustrate how central banks have increasingly moved from judgment-based assessments of tail risk toward transparent and replicable econometric measurement.

\paragraph{Summary of Federal Reserve practice.} Risk has been monitored at the Fed through two main quantitative approaches. Scenarios have appeared in the Tealbook since 1995. Predictive densities have been elicited through the SPF (since 1968), built by the NY Fed Blackbook by weighting scenarios judgmentally (since about 2000), surveyed via SEPIA (2017), and produced statistically by the Board's TVMR exhibit using financial conditions (2017). Crucially, the NY Fed's approach of building densities \emph{from} scenarios is the essential intuition the Scenario Synthesis formalizes. 

There is also a third channel, not reviewed separately above because it is not a quantitative tool in the same sense: the staff's judgmental \emph{Assessment of Risks}. This assessment provides a qualitative characterization of the balance of risks and uncertainty around the baseline and often carries substantial weight in policy deliberations. It may draw on scenarios, predictive densities, and other statistical exhibits, but it is not reducible to any one of them. In July 2019, for example, staff judged risks to be tilted to the downside even though the one-year-ahead time-varying-risk estimates were ``not unusually wide or skewed'' \citep[][p.~69]{Jul2019Tealbook}. Scenario Synthesis can be viewed as a structured way to make this judgmental assessment explicit, quantitative, and reproducible.

This historical reconstruction shows that the Federal Reserve has developed a rich but fragmented toolkit for assessing macroeconomic risk. We next place these practices in a broader international context.

%
%
\section{Risk Analysis at Other Central Banks and Institutions}\label{sec:other_cbs}

The Federal Reserve's experience is part of a broader global effort. Other central banks, international institutions, and private firms have developed their own approaches---sometimes independently, sometimes influenced by the Fed. We organize the discussion around the three building blocks introduced above: baseline projections, scenarios, and predictive densities.

\subsection{Baselines Across Institutions}\label{sec:baselines_other}

Across central banks, baseline projections are typically produced with a combination of structural and reduced-form models. We briefly illustrate this with two prominent examples.

\paragraph{European Central Bank.}
The ECB's baseline projections are staff projections supported by a suite of models. Two central components are ECB-Base \citep{ECB-BASE}, a large semi-structural model for the euro area, and ECB-MC \citep{ECB-MC}, a multi-country semi-structural model covering euro-area member countries; both are designed in the same broad tradition as FRB/US. The third central component is the New Area-Wide Model~II (NAWM~II), a micro-founded open-economy DSGE model estimated on euro-area data using Bayesian methods \citep{christoffel2008,coenen2018}. The ECB also maintains a range of satellite models---including small DSGE models, small semi-structural models, and time-series models---that provide cross-checks on the baseline projection \citep[see][for a description of satellite models]{ECB-MacroModels}.

\paragraph{Bank of England.}
The Bank of England's baseline model is COMPASS, a medium-scale open-economy New Keynesian DSGE model estimated on UK data \citep{burgess2013}. Adopted in 2011, it was designed as a central organizing model within a forecasting platform that also includes more than forty supplementary models. A medium-scale Bayesian VAR serves as its reduced-form companion \citep{Domitetal2019}, matching the DSGE's coverage while being more robust to structural misspecification. In practice, however, the Bank's forecast infrastructure relies heavily on judgment and supplementary models, so the published baseline is best understood---as at the Federal Reserve---as a judgmental staff product rather than the output of a single model.\footnote{In his independent review, \citet{bernanke2024} recommended replacing or substantially revamping COMPASS, citing limitations in its monetary-transmission channels. Consistent with this modernization agenda, the Bank of England released an updated medium-scale DSGE model for the UK economy in 2025 \citep{albuquerque2025dsgeuk}.}


\subsection{Scenarios Across Institutions}\label{sec:scenarios_other}

The Federal Reserve Board has the longest continuous tradition of systematic scenario analysis, producing scenarios in the Tealbook since at least 1995. Elsewhere, scenario analysis was historically reserved for exceptional circumstances or internal stress tests rather than regular communication. That changed after the Global Financial Crisis, which exposed the limits of single-baseline forecasting under deep uncertainty. Since then, scenarios have spread widely, though formalization, publication frequency, and integration with probabilistic tools vary considerably.

\paragraph{European Central Bank.}
Before the pandemic, the ECB used scenarios only occasionally. COVID-19 was a turning point: in June 2020, staff first built three pandemic scenarios---mild, severe, and very severe---and only then chose which would become the published baseline, inverting the usual logic and treating the baseline as one element of a broader scenario space. Scenarios have since become regular features of ECB projections and communication, used extensively during the Russia-Ukraine energy shock and, more recently, for U.S.\ tariff risks and Middle East geopolitical risks. In each case, they structured deliberations around defined contingencies and communicated uncertainty to the public.

\paragraph{Bank of England.}
The Bank historically communicated uncertainty through fan charts: probability distributions around the central projection, calibrated from forecast errors and judgment, that became a widely imitated template. Following the \citet{bernanke2024} review, which criticized fan charts as difficult to interpret and weak in conveying the narratives underlying risk assessments, the Bank began publishing explicit scenarios in November 2025; in the April 2026 \emph{Monetary Policy Report}, amid exceptional uncertainty about the Middle East energy shock, it even replaced the baseline with three scenarios.\footnote{The May 2020 projection was also presented as a ``scenario,'' but then, between May 2020 and November 2025, the Bank did not publish any other scenario.} This marks a shift from a statistical summary of forecast risk to a narrative account of the paths the Monetary Policy Committee (MPC) considers most relevant.\footnote{The shift is one of emphasis, not abandonment: fan charts remain in the public databank, now calibrated solely on past forecast errors.}

\paragraph{Other central banks.}
In April 2025, the Bank of Canada departed from its conventional single-baseline forecast, communicating its outlook \emph{exclusively} through two scenarios without assigning relative probabilities, on the grounds that deep uncertainty about U.S.\ trade policy made any baseline spuriously precise. The Riksbank has published scenarios regularly since 2023 and is unusual in reporting an explicit interest-rate path for each, thereby clarifying its reaction function under alternative outcomes. The Central Bank of Armenia has gone further, adopting a purely scenario-based framework and replacing the single baseline altogether.

\paragraph{International institutions.}
The IMF's \emph{World Economic Outlook} has published scenarios around its baseline forecast for nearly a decade, covering risks such as financial stress, trade tensions, commodity-price shocks, and geopolitical disruptions. These scenarios are typically presented as deviations from the global baseline, built with the IMF's global macroeconomic model, and accompanied by narratives describing the shock assumptions and transmission channels. The OECD similarly publishes risk assessments and alternative scenarios in its \emph{Economic Outlook}. Both institutions have also developed Growth-at-Risk frameworks, so scenario-based and distributional approaches increasingly coexist within the same institution, closely mirroring the parallel structure observed in the Tealbook.

\paragraph{Private institutions.}
Private forecasting firms---Oxford Economics, Moody's Analytics, S\&P Global, and others---have long offered scenario analysis as a commercial product. Their scenarios provide probability-weighted alternative paths for the global economy, individual countries, and sectors, and are used for portfolio stress testing, strategic planning, and risk management. Unlike most central-bank scenarios, they often carry explicit probability weights, reflecting client demand for a single risk-adjusted metric.

\paragraph{Supervisory bank stress testing.}
A separate tradition applies scenario-based stress testing to financial institutions. Originating at the IMF in the late 1990s, this practice subjects bank balance sheets to adverse macroeconomic scenarios---severe recessions, asset-price declines, and credit-spread spikes---and assesses the resulting capital shortfalls \citep[see][]{Adrian2020}. After the Global Financial Crisis, it became a regulatory requirement in most major jurisdictions. This tradition differs from central-bank forecasting because it focuses on institution-level solvency rather than economy-wide dynamics, but it rests on the same logic: concrete, narratively coherent scenarios are often more decision-relevant than abstract probability distributions alone.

\subsection{Predictive Densities Across Institutions}\label{sec:densities_other}

\paragraph{Judgmental approaches.}
The Bank of England was the first central bank to communicate macroeconomic risk explicitly through predictive densities. Its fan charts, introduced in the \emph{Inflation Report} in 1996, reported quantiles of the predictive distribution of inflation---and later GDP growth and unemployment---using models and judgment \citep{BankofEngland1998}. The European Central Bank has employed a different judgmental approach through the \emph{Quantitative Risk Assessment} (QRA), an internal survey-based tool that summarizes ECB staff views on the main risk events surrounding the baseline, including their upside or downside nature and their implications for uncertainty and skewness \citep{ECBWorkstream1_2025}. The QRA is not a stand-alone survey reference like the SPF, but one element of the internal risk-assessment process. 

\paragraph{Survey-based approaches.}
The ECB has operated a Survey of Professional Forecasters since its inception, modeled in part on the Philadelphia Fed SPF. It asks respondents for probabilistic assessments of GDP growth, inflation, and unemployment, providing survey-based predictive distributions that complement staff risk assessments. The ECB also draws on other surveys, including the Survey of Monetary Analysts and the Consumer Expectations Survey, which contain probabilistic information relevant for assessing expectations and risks. Similar surveys are conducted by several other central banks, including the Bank of Japan and the Bank of Canada.

\paragraph{Statistical approaches.}
The ECB also reports uncertainty ranges based on past projection errors, a backward-looking benchmark that complements forward-looking density tools. The most important recent development in quantitative risk analysis has been the worldwide adoption of Growth-at-Risk (GaR) models. The IMF introduced GaR in the October 2017 \emph{Global Financial Stability Report} \citep{IMF2017}, applying it across 21 economies; \citet{Adrianetal2022} later developed its term structure, documenting that loose financial conditions reduce near-term downside risk but increase it at longer horizons. GaR-based forecasts now complement risk toolkits at the ECB \citep{FIGUERES2020109126,ECBWorkstream1_2025,lenza2023density}, the Bank of England \citep{aikman2019credit,LloydMantoanManuel2022,lloyd2024foreign,Anesti2023,GaR_BoE2024}, the Banque de France \citep{BdFGaR2022,FerraraMoglianiSahuc2022,Lhuissier2022}, and several other central banks.\footnote{See, among others, applications at the Bank of Italy \citep{BoIGaR2019}, the Central Bank of Ireland \citep{CBIGaR2021}, the Bundesbank \citep{BundesbankGaR2025}, Banco de Espa\~na \citep{Galan2020}, the Bank of Japan \citep{BoJInflationAtRisk2022}, and Norges Bank \citep{Boweetal2023}.}

Evaluating these tools against ease of communication, narrative content, forward-looking nature, and probability assessment, \citet{ECBWorkstream1_2025} find that statistical models score high on probability assessment and forward-looking content but low on narrative---the mirror image of scenarios. Across institutions, then, the toolkit for risk analysis is broad but fragmented, with different tools emphasizing different dimensions of risk. We now turn to the disconnect this produces.

%
%

\section{The Disconnect Between Scenarios and Predictive Densities}\label{sec:disconnect}
Across institutions, scenarios and predictive densities coexist within the same forecasting and policy process, yet they are rarely embedded in a common framework. They are often produced using different models and information sets: scenarios deliver narratives without probabilities, while predictive densities deliver probabilistic risk assessments with limited economic interpretation.

The disconnect persists because their strengths are complementary. Scenarios are rich in narrative and broad in scope---large structural models, causal stories, joint forecasts---but limited in their treatment of risk: linear or linearized models make uncertainty largely symmetric and time-invariant, with little room for fat tails, skewness, or state-dependent amplification. Predictive densities are richer in risk---capturing asymmetry, nonlinearity, and time-varying tails---but reduced-form and narrower in scope, offering less on mechanisms and typically covering fewer variables.

Scenario analysis also serves distinct purposes, which shape what a scenario set should contain. Some scenarios communicate the balance of risks around the outlook, as in the Tealbook's ``Risks and Uncertainty'' chapter. Others communicate the central bank's reaction function, as \citet{Bernanke2025} and \citet{Gargaetal2025} emphasize, making the policy path itself an object of interest. Supervisory stress tests, by contrast, deliberately probe the tail, whereas monetary-policy scenarios typically capture large but not extreme risks. Scenario Synthesis is agnostic across these uses: it evaluates any scenario set against the reference density that encodes the relevant notion of risk.

The balance is shifting. As macroeconomic risks have intensified---through the Global Financial Crisis, the COVID-19 pandemic, the Russia-Ukraine war, and tariff shocks---institutions that previously emphasized predictive densities have increasingly turned to scenarios. The ECB's expansion since 2020, the Bank of England's published scenarios in 2025, and \citeauthor{Bernanke2025}'s proposal that the Fed publish scenarios after each FOMC meeting all point in this direction. Surveying 25 central banks, \citet{BellChavazHofmannReesRottner2026} find that the share using scenario analysis rose from about 25\% before the pandemic to about 45\% in 2025.

As central banks increasingly use scenarios and predictive densities side by side, the risk of conflicting signals grows. The natural question is how to combine the narrative richness of scenarios with the quantitative discipline of predictive densities. The Scenario Synthesis does so by choosing weights on the baseline and alternative scenarios so that their mixture best approximates a reference predictive distribution under a concordance criterion. The resulting Synthesis links the structural and communicative strengths of scenarios with the probabilistic rigor of densities.

The framework also connects to the debate on risk communication. \citet{bernanke2024} recommended de-emphasizing fan charts in favor of qualitative risk descriptions and narrative scenarios. Scenario Synthesis bridges these approaches by preserving scenario narratives while attaching transparent weights, and is complementary to approaches that attach policy-rate paths to scenarios \citep{LaxtonIgityanMkhatrishvili2025,Bernanke2025,Gargaetal2025}.

A central implication is that weights depend on the reference density. Different predictive densities---that is, different assessments of risk---imply different scenario weights. When downside risk is small, adverse scenarios receive little weight and the set may appear adequate; when downside risk is large, adverse scenarios receive greater weight and shortcomings become visible. Sections~\ref{sec:2007} and~\ref{sec:2018_uni} show that this dependency matters in practice.

The Scenario Synthesis framework is beginning to enter policy practice. At the Banque de France, \citet{Lhuissier2026} applies it to the Eurosystem's June 2025 projections, using the ECB Survey of Professional Forecasters as the reference distribution to construct a synthesis-based optimal monetary policy path. At the Bank of England, a cross-divisional team applies the framework to the three scenarios in the April 2026 \emph{Monetary Policy Report}. At the three-year horizon, they find that the Synthesis captures about 87\% of the inflation risk in the Decision Maker Panel survey and spans roughly the 25th--75th percentiles of the one-year-ahead market-implied Bank Rate distribution. In Section~\ref{sec:empirics}, we apply the framework to Federal Reserve Tealbook scenarios in December 2007 and December 2018.

%
%
\section{The Framework}\label{sec:toolkit}

The economy is complex, nonstationary, and possibly indeterminate. Its dynamics depend on observable variables y (GDP, inflation, unemployment, etc.) and on a large set of largely hidden conditions z (financial frictions, expectations, global conditions, regime indicators, etc.) that no single model can fully capture. Central bank forecasting is therefore an inference problem under deep uncertainty: policymakers cannot agree on a single representation of the economy, nor do they observe the full economic state directly; they observe only aggregates such as GDP, inflation, unemployment, spreads, and surveys, together with a finite historical sample.

As we have documented in Sections~\ref{sec:fed_history} and \ref{sec:other_cbs}, central banks confront and communicate this uncertainty using a multi-layered forecasting apparatus. The toolkit typically includes a \emph{baseline projection}, $p_0(y)=p(y\mid A_0,M_0)$, that delivers a coherent economic narrative under a set of conditioning assumptions $A_0$ describing ``normal'' states, and the baseline model $M_0$; a collection of \emph{alternative scenarios}, $p_j(y)=p(y\mid A_j,M_j)$, $j=1,\ldots,J$, that examine specific risks under alternative conditioning assumptions $A_j$, and model $M_j$ chosen to capture the mechanisms deemed relevant for the risk under consideration; and a \emph{reference predictive density}, $p(y)$, that aims to characterize the full range of possible outcomes regardless of which conditioning assumptions or model happen to be relevant via flexible statistical methods and/or expert judgment. We discuss each in turn before showing, in the next section, how the Scenario Synthesis links them.

\subsection{Baseline projection}
As documented in Sections~\ref{sec:fed_scenarios} and \ref{sec:baselines_other}, most major central banks maintain a large-scale structural model that plays the role of the baseline model ($M_0$), providing a coherent economic narrative for the institution's point forecast. Producing the baseline forecast combines $M_0$ with baseline assumptions $A_0$ about variables, shocks, and mechanisms treated as exogenous: paths for global demand, foreign output, commodity prices, fiscal policy, and shock distributions. The baseline predictive density is thus
\vspace{-12pt}\[ p_0(y) \equiv p(y \mid A_0, M_0).\vspace{-12pt}\]

Baseline models are typically large macroeconometric systems, with hundreds of equations covering consumption, investment, housing, labor markets, prices, fiscal and monetary policy, and international linkages. They combine micro-founded components with reduced-form elements, such as empirical Phillips curves, and build in several practical approximations. First, they rely on linear or log-linear approximations around a ``normal'' baseline, limiting their ability to capture strong nonlinearities, regime shifts, and binding constraints. Second, shock variances are typically fixed, so risk is treated as time-invariant and effects are often symmetric. Third, financial frictions---runs, fire sales, and balance-sheet interactions---are represented only partially. Fourth, they are optimized for ``normal'' business-cycle environments rather than stress episodes or structural breaks. Even so, they remain central: they provide a shared normal-times benchmark, support causal narratives, and organize communication and decomposition exercises.

\subsection{Alternative scenarios}

Central banks complement the baseline with alternative scenarios designed to explore specific risks or alternative mechanisms. A scenario is defined by a particular combination of assumptions $A_j$ and model $M_j$:
\vspace{-12pt}\[ p_j(y) = p(y\mid A_j, M_j), \qquad j=1,\ldots,J. \vspace{-12pt}\]
Here $A_j \neq A_0$, while $M_j$ may differ from $M_0$. 

The assumptions $A_j$ are understood broadly. They include paths or ranges for exogenous variables, shock distributions, qualitative states, and the assumed monetary policy reaction function. Examples include oil prices are elevated,'' financial stress is high,'' or ``foreign demand is weak.'' These assumptions define regions of the state space rather than points, so the associated conditional densities can receive non-negligible weight even when the scenario is communicated as a central path. The policy assumption also matters for interpretation. A scenario is a conditional density given an assumed policy response, so the synthesis weights are conditional on those policy assumptions as well; see Sections~\ref{sec:economics} and \ref{sec:policy}.

Because $M_0$ is a local approximation around normal conditions, assumptions and model structure are often intertwined. In stressed or nonlinear environments, the local approximation may become unreliable and a different model $M_j$ may be needed: a credit-accelerator block for severe financial stress, a nonlinear Phillips curve for large supply shocks. By contrast, $M_0$ can handle a mild foreign-growth slowdown. The case $M_j \neq M_0$ is not confined to extreme stress. Some first-order risks concern structural change---for example, greater intrinsic persistence of wage- and price-setting after high inflation---and are naturally captured by a model $M_j$ that better reflects the mechanism. The principle is to match the model to the assumptions under consideration.

Scenarios examine regions of the state space relevant for policy but not central enough for $A_0$. They are not competing forecasts but conditional ``if--then'' explorations.\footnote{As \citet{crumpetal2025} emphasize, conditional forecasting links the central-bank scenario tradition to structural analysis based on estimated impulse responses.} They help policymakers reason under deep uncertainty.\footnote{We use ``deep uncertainty'' in the sense of the operations-research literature \citep{WalkerLempertKwakkel2013}: situations in which decision-makers cannot agree on the model, the relevant outcomes, or the probability distributions over inputs. We develop this notion and its implications for policy in Section~\ref{sec:policy}.}

\subsection{Reference predictive density}

The reference predictive density $p(y)$ serves a distinct purpose from both the baseline and the scenarios: it approximates the unconditional predictive distribution as accurately as possible, without necessarily providing a structural narrative. Researchers typically build it with flexible reduced-form methods designed for predictive accuracy---high-dimensional quantile regressions, machine learning, Bayesian model averaging, time-varying-parameter models, and Growth-at-Risk frameworks. Its strengths are out-of-sample accuracy, nonlinearities, time-varying risk, and large information sets; its weaknesses are limited structural interpretation and communicability.\footnote{\citet{Schroder2025} addresses this limitation by decomposing inflation-at-risk densities into underlying risk drivers.}

Two points bear on how the synthesis should be read. First, the framework is agnostic about which reference is used: $p(y)$ can be any density the user regards as a good unconditional predictive distribution---a Growth-at-Risk model, a time-varying-parameter model, a survey-based density, or a judgmental assessment. If a user believes a statistical reference understates nonlinear ``dark-corner'' tail risks---for example, convex-Phillips-curve or credibility-loss dynamics---then the user should supply a reference that does not. A scenario designed to capture such a risk will receive low weight against a reference that underweights the tail, but this signals a limitation of the reference rather than an implausible scenario---exactly the type of diagnostic the synthesis is meant to surface.

Second, when policymakers disagree about the reference, the response is not to force a single choice but to report the synthesis under each candidate, treating the spread of weights as a robustness band. This is the approach we take in Section~\ref{sec:empirics}.

%
%

\section{The Scenario Synthesis}\label{sec:scenario_synthesis}

The baseline model $M_0$ provides interpretability and causal structure, but delivers an incomplete characterization of the range of outcome omitting macro-financial linkages, nonlinearities, and time-varying risk that shape the tails. Scenarios complement it by articulating ``what-if'' risks not embedded in the baseline. The reference density $p(y)$ delivers a richer characterization of the distribution of outcomes, but at the cost of interpretability: we know that elevated credit growth fattens the left tail, but the mechanisms remain debated. 

Scenario Synthesis connects what we have learned empirically (the reference) with what we understand structurally (the baseline and scenarios). We present the framework here and refer to \citet{AGLW2025,AGLW2026stats} for  more statistical details. The synthesis explores and quantifies how well  the reference predictive distribution $p(y)$ can be approximated by a statistical mixture of the model-based conditional views:
\begin{equation}
    \fya = \sum_{j=0}^J \alpha_j \, p_j(y), \qquad \alpha_j \ge 0, \quad \sum_{j=0}^J \alpha_j = 1,
\end{equation}
where the weights $\balpha$ quantify the contribution of each conditional view to the synthesis.
 
This mixture scenario synthesis resembles predictions using traditional Bayesian Model Averaging (BMA) under model uncertainty. In BMA, the $p_j(y)$ are predictions from a set of assumed models and the $\alpha_j$ are model probabilities. These probabilities are based on past data that have informed the models' relative predictive performance. The more general framework of Bayesian predictive synthesis (BPS: \citealp{TallmanWest2023,JohnsonWest2024}) allows the $\alpha_j$ to be specified otherwise, while maintaining the traditional mixture form. Note that there is no notion that $\fya$ is the data-generating mechanism: the $\alpha_j$ are wholly relative across the model set, with models weighted relative to one another based solely on predictive (and, in some cases, decision) outcomes. There is no assumption that there is a \lq\lq true model'' within the set~\citep[][sect.~12.2]{WestHarrison.YellowBook2ndEdn.1997}.  

Scenario Synthesis adopts the mixture form inspired by BPS/BMA, but the interpretation differs. Each scenario is a distinct conditional distribution $p(y\mid A_j,M_j)$: a complementary exploration of an alternative state of the world, not a competing unconditional forecast. The synthesis weights are chosen so that $\fya$ best approximates the reference density $p(y)$. While the weights have the mathematical form of probabilities on the finite scenario set, they are not generally interpretable as subjective or data-based probabilities that the scenarios will occur. Rather, they are diagnostic importance measures relative to both the scenario set and the reference being matched: a high weight means that a scenario better approximates the reference than the other scenarios considered, not that the scenario is likely in an absolute sense.

There is one special case in which the weights have a direct probabilistic interpretation. Suppose that the baseline, scenarios, and reference density are all generated from the same model $M$, so that $M_j=M$ for all $j$, and that the conditioning assumptions $\{A_j\}$ form a mutually exclusive and exhaustive partition of the state space. Then
\vspace{-12pt}\[
p(y\mid M)=\sum_{j=0}^J p(y\mid A_j,M)p(A_j\mid M),
\vspace{-12pt}\]
so that the synthesis weights coincide with the marginal probabilities of the conditioning assumptions, $\alpha_j^*=p(A_j\mid M)$. Outside this case, we refer to the $\alpha_j^*$ as weights rather than probabilities.

\subsection{Optimizing the weights}\label{sec:optimizing_the_weights}

The synthesis weight vector $\balpha=(\alpha_0,\ldots,\alpha_J)'$ is chosen such that the mixture is as ``close'' as possible to the reference. To formalize closeness, we use a concordance measure from statistical classification: the expected misclassification rate (EMR: see~\citealp{AGLW2025,AGLW2026stats} for statistical background). This is given by
\vspace{-6pt}\[\emr(\balpha) =  \int_y  \frac{\fya\py}{\{\fya+\py\}}dy.\vspace{-6pt}\]
Higher values of $\emr(\balpha)$ indicate greater probabilistic concordance between the mixture $\fya$ and the reference $\py$. Since $\emr(\balpha)\in(0,0.5]$, values close to $0.5$ indicate high concordance.

We regularize the optimization by placing a weak Dirichlet prior on the weights, setting $\epsilon=0.05$ as in \citep{AGLW2025}. Specifically, with $\balpha\sim \mathrm{Dir}(\mathbf{1}(1+\epsilon))$, the posterior mode solves
\begin{equation}
  \{\alpha_j^*\}_{j=0}^J
  =
  \argmax_{\substack{\alpha_j>0,\ \alpha_0 \ge \alpha_j \\
  \sum_{j=0}^J \alpha_j = 1}}
  \left[
  \log\{\emr(\balpha)\}
  +
  \epsilon\sum_{j=0}^J \log(\alpha_j)
  \right]. 
\end{equation}
The Dirichlet term is not needed to define concordance; it regularizes the weights, avoiding unstable zero weights when scenarios are nearly redundant.  The restriction $\alpha_0 \ge \alpha_j$ is likewise not required mathematically; we impose it to reflect the institutional role of the baseline as the central projection around which scenarios are constructed. It prevents any single alternative scenario from receiving more weight than the baseline while still allowing the scenario set as a whole to dominate the baseline when the reference calls for it.

The residual divergence
\begin{equation}
  \demr = 0.5-\emr(\balpha^*)
\end{equation}
measures how far the optimally weighted scenario set is from spanning the reference distribution. A value of $\demr$ close to zero indicates that the scenario set captures the reference well, whereas larger values reflect missing or redundant scenarios. The diagnostic is asymmetric: a large $\demr$ indicates that the scenario set does not span the reference, while a small $\demr$ establishes numerical fit but not necessarily economic coverage. For the latter interpretation, the scenarios must represent meaningful regions of the state space.

%
%
\section{The Scenario Synthesis at Work}\label{sec:empirics}

\subsection{December 2007}\label{sec:2007}

We apply the Scenario Synthesis to the alternative scenarios presented in the December 2007 Greenbook \citep{Dec2007Tealbook}, one year before the Great Recession. The episode is a natural application because two predictive densities with sharply different risk assessments were available: the SPF consensus and the NY Fed Blackbook's judgmental density (BB).\footnote{Real-time Outlook-at-Risk estimates do not exist for 2007, as the underlying quantile-regression methodology postdates it; applying a method introduced a decade later would be inappropriate. The Blackbook and SPF densities, by contrast, reflect contemporaneous assessments, avoiding hindsight bias.} Holding the Tealbook scenario set fixed, we vary the reference distribution---a controlled comparison isolating how alternative risk assessments shape the implied weights and diagnostics.

Although our methodology applies to multiple variables and horizons, we focus on one-year-ahead GDP growth in this section to clarify the mechanics. In Section~\ref{sec:Dec2018_GDPPCE}, we extend the analysis to the joint distribution of GDP growth and core PCE inflation. \ref{sec:scenario_synthesis_implementation} summarizes the main computational steps.

In the empirical applications, we report both the EMR and the Effective Sample Size statistic (ESS). The two are monotonically related. We report both because the EMR is the optimization objective, while the ESS is easier to read; equivalently, $100-\mathrm{ESS}$ provides an intuitive measure of residual non-overlap. Since the EMR flattens at high concordance, small EMR differences can correspond to economically meaningful differences in spanning.

\subsubsection{Scenario Set and Modeling Framework}\label{sec:2007_scenarios}

The December 2007 Greenbook features seven domestic alternative scenarios.\footnote{The December 2007 Greenbook also included a separate set of international alternative simulations; presenting domestic and international scenarios separately was standard practice until 2008 \citep{HerbstKonzemScofield2026}. We do not incorporate the international simulations into our analysis. We flag them, however, because one included a downside risk to the U.S.\ outlook. This qualifies, but does not overturn, our conclusion that the domestic scenario set understated downside risk to GDP growth.} The ``Greater Housing Correction'' scenario ($\scn_1$) assumes a more severe deterioration in the housing market than in the Baseline. The ``Credit Crunch'' scenario ($\scn_2$) examines a situation in which financial market turbulence leads to a sharp tightening of credit conditions, significantly restricting lending to businesses and households.

Scenarios $\scn_3$ and $\scn_4$ represent upside demand risks. The ``Stronger Domestic Demand'' scenario ($\scn_3$) considers the possibility that financial stress exerts less drag on spending than in the Baseline. Building on this, the ``Stronger Domestic Demand with Better Export Performance'' scenario ($\scn_4$) adds the assumption of stronger export growth.

The ``More Room to Grow'' scenario ($\scn_5$) assumes faster potential output growth, while the ``Greater Cost Pressures'' scenario ($\scn_6$) assumes that firms raise prices more aggressively in response to rising costs. Finally, the ``Market-Based FFR'' scenario ($\scn_7$) assumes that the federal funds rate evolves according to the path implied by futures markets.

All scenarios are constructed using the same baseline model (FRB/US). In all but one, monetary policy responds via an estimated Taylor rule; the exception is the ``Market-Based FFR'' scenario ($\scn_7$), in which the federal funds rate follows the path implied by futures markets. In the notation of Section~\ref{sec:toolkit}, this implies that $M_j = M_0$ for all $j$, so that variation across scenarios arises from the conditioning assumptions $A_j$---including, for $\scn_7$, the assumed policy path. The scenarios are therefore intended to capture distinct economic mechanisms through alternative assumptions, although in practice some overlap remains.

\begin{wraptable}{r}{.6\textwidth} \setstretch{1.0} \vspace{-12pt}
\caption{Baseline and Alternative Scenarios} \label{tab:Baseline2007}  \small \centering 
\footnotesize{Dec. 2007 Tealbook --- One-year ahead GDP growth projection}\\[2pt]
\begin{tabular}{L{.005\textwidth} L{.025\textwidth} L{.325\textwidth}R{.075\textwidth}R{.075\textwidth}R{.075\textwidth}R{.005\textwidth}} \hline\hline     
  &{$j$} &   {Scenario $\scn_j$}      &   {P15} &   {P50} &   {P85}      &\\ \hline
  &  0  &   Baseline                        &    0.1    &   1.3     &  2.5 &\\
  &  1  &  Greater housing correction       &       &   0.9     &    &\\
  &  2  &  Credit crunch                    &       &  $-$0.4   &    &\\
  &  3  &  Stronger domestic demand         &       &   1.7     &    &\\
  &  \grigio 4  &  \grigio{Better export performance}        &       &   \grigio{1.9}     &    &\\
  &  5  &  More room to grow                &       &   1.9     &        \\    
  &  6  &  Greater cost pressure            &       &   1.2     &    &\\
  &  7  &  Market-based FFR                 &       &   1.6     &    &\\\hline     
\end{tabular}

\begin{tabular}{p{.585\textwidth}} \scriptsize
\textsc{Note:} The baseline projection for 2008 is reported on page~I-21; scenario projections appear on page~I-17. Scenario values for 2008 are obtained by averaging 2008:H1 and 2008:H2. P50 is the point forecast (Baseline and Scenarios), while P15 and P85 are the bounds of the 70\% interval. The Tealbook reports only the point forecast of the alternative scenarios, so their P15 and P85 are left blank.
\end{tabular}\vspace{-12pt}
\end{wraptable}
We exclude $\scn_4$ (``Better Export Performance'') from the analysis because, as shown in Table~\ref{tab:Baseline2007}, its one-year-ahead GDP growth projection is identical to that of $\scn_5$ (``More Room to Grow''). The two scenarios differ along other dimensions---inflation, unemployment rate, and the federal funds rate---but they are indistinguishable in the univariate analysis focused on GDP growth. We therefore drop $\scn_4$ under the exclusion criterion used throughout: when two scenarios imply the same projection in the variable space under consideration, one is redundant for the optimization and the individual weights are not uniquely identified (Section~\ref{sec:optimizing_the_weights}). Removing $\scn_4$ also leaves a set with non-overlapping conditioning assumptions in the univariate space, since $\scn_4$ combines stronger domestic demand with better export performance and therefore nests $\scn_3$.

\subsubsection{Reference and Baseline Distributions}\label{sec:2007_Reference}
We now present the two references and compare them with the Baseline. We use the \citet{crumpetal2025} Large BVAR to convert the Blackbook and SPF forecasts for annual-average GDP growth into quantile forecasts for Q4/Q4 growth, as detailed in \ref{sec:annual_2_q4q4}.\footnote{Because the Blackbook reports no numeric percentiles, we extract the histogram values using Optical Character Recognition; see \ref{sec:OpticalCharacterRecognition}.} We then fit a \skt density \citep{AzzaliniCapitanio2003} to the resulting quantiles, choosing its four parameters to minimize the squared distance between target and fitted quantiles, following \citet{Adrianetal2019}.

For the Baseline, the Tealbook reports a point forecast and a 70\% interval (Table~\ref{tab:Baseline2007}). We treat the point forecast as the median and the interval bounds as the 15th and 85th percentiles, and fit a \skt with location at the point forecast, zero skewness, and degrees of freedom fixed at 50.

\begin{figure}[h]\caption{Reference -- Baseline -- Scenarios}\label{fig:RBS_GDP_2007}
\centering
\textsc{\small 2008 Q4/Q4 GDP growth -- Dec. 2007 Tealbook}\\ \smallskip
\begin{tabular}{C{.5\textwidth}C{.5\textwidth}}
\footnotesize \sc Blackbook Reference & \footnotesize  \sc SPF Reference \\ 
\includegraphics[width=.49\textwidth]{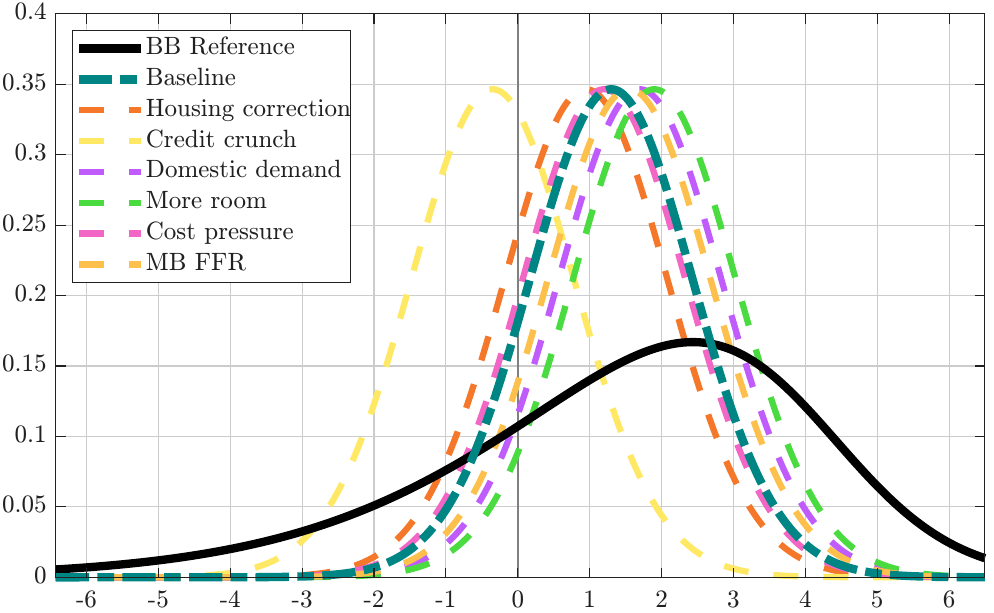} &
\includegraphics[width=.49\textwidth]{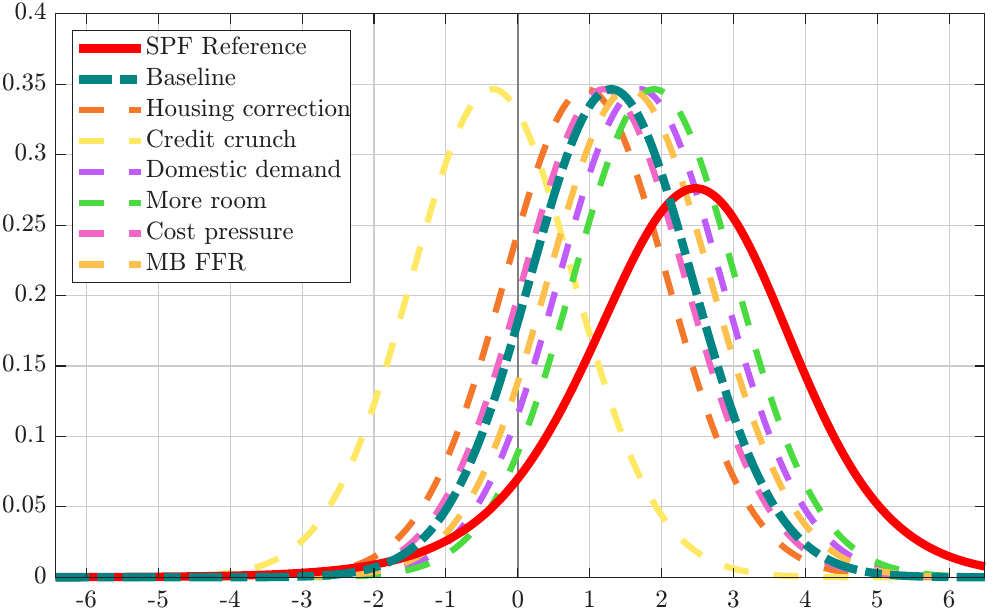}\\
\end{tabular}

\vspace{-3pt}
\begin{tabular}{p{\textwidth}}\scriptsize
\textsc{Note:} In the left panel, the solid black line denotes the Blackbook Reference p.d.f. In the right panel, the solid red line denotes the SPF Reference p.d.f. In both panels, the teal dash-dotted line denotes the Baseline p.d.f., while the dashed lines show the scenario p.d.f.s.
\end{tabular}
\end{figure}

Figure~\ref{fig:RBS_GDP_2007} plots the Blackbook reference distribution (black), the SPF reference distribution (red), the Baseline (teal dash-dotted), and the alternative scenarios (shown as point forecasts in the left panel and as predictive densities in the right panel). The Baseline differs substantially from the BB Reference in both location and shape. By contrast, 
\begin{wraptable}{r}{.6\textwidth} \setstretch{1} 
\caption{Skew-$t$ parameters} \label{tab:RB_SKTparam2007}\centering\small
\begin{tabular}{L{.025\textwidth}L{.175\textwidth}C{.1\textwidth}C{.1\textwidth}C{.1\textwidth}C{.1\textwidth}}
\hline \hline
&& lc & sc & sk & df \\\hline
&BB Reference  & 4.2 & 3.6 & -2.3 & 14.3 \\
&SPF Reference & 3.0 & 1.5 & -0.6 & ~8.8 \\
&Baseline      & 1.3 & 1.1 & ~0.0 & 50.0 \\\hline
\end{tabular} 

\begin{tabular}{p{.6\textwidth}} \scriptsize
\textsc{Note:} This table shows the \skt fitted parameters for the Blackbook and SPF References and the Baseline. The parameters reported in the table correspond to the location (lc), scale (sc), skewness (sk), and degrees of freedom (df).
\end{tabular}
\vspace{-12pt}
\end{wraptable}
it primarily diverges from the SPF Reference in terms of location; see also the estimated \skt parameters reported in Table~\ref{tab:RB_SKTparam2007}. Visually, the figure already suggests that the scenario set is too concentrated around the Baseline to span either reference distribution well.

The two references convey sharply different risk assessments. The BB Reference is more dispersed and strongly negatively skewed, with substantial left-tail mass: it assigns 25\% probability to negative GDP growth, consistent with FRBNY staff concern about financial conditions. The FRB staff expressed related concerns in the Tealbook, replacing the previous ``Greater housing correction with larger fallout'' scenario with the more severe ``Credit Crunch'' scenario. By contrast, the SPF Reference is more concentrated, closer to symmetric, and shifted to the right, assigning only 7\% probability to negative growth. The two references thus emphasize different regions of the outcome space---downside tail risks versus relatively favorable outcomes---providing a natural setting to assess whether the Tealbook scenario set can match alternative risk assessments.

\subsubsection{Scenario Synthesis}\label{sec:2007_NYFed}

We now evaluate the Tealbook scenario set against each reference. Because the December 2007 Tealbook reports scenarios only as point forecasts, we construct scenario densities by shifting the location of the Baseline density to match each point forecast while keeping scale, skewness, and degrees of freedom fixed (see also Section~\ref{sec:ScenarioDistributions}).

\begin{figure}[h]\caption{Scenario Synthesis --- Probability density functions}\label{fig:ScenarioSynthesis2007GDP}
\centering
\textsc{\small 2008 Q4/Q4 GDP growth -- Dec. 2007 Tealbook}\\ \smallskip
\begin{tabular}{C{.5\textwidth}C{.5\textwidth}}
\footnotesize \sc Blackbook Reference & \footnotesize  \sc SPF Reference \\ 
\includegraphics[width=.49\textwidth]{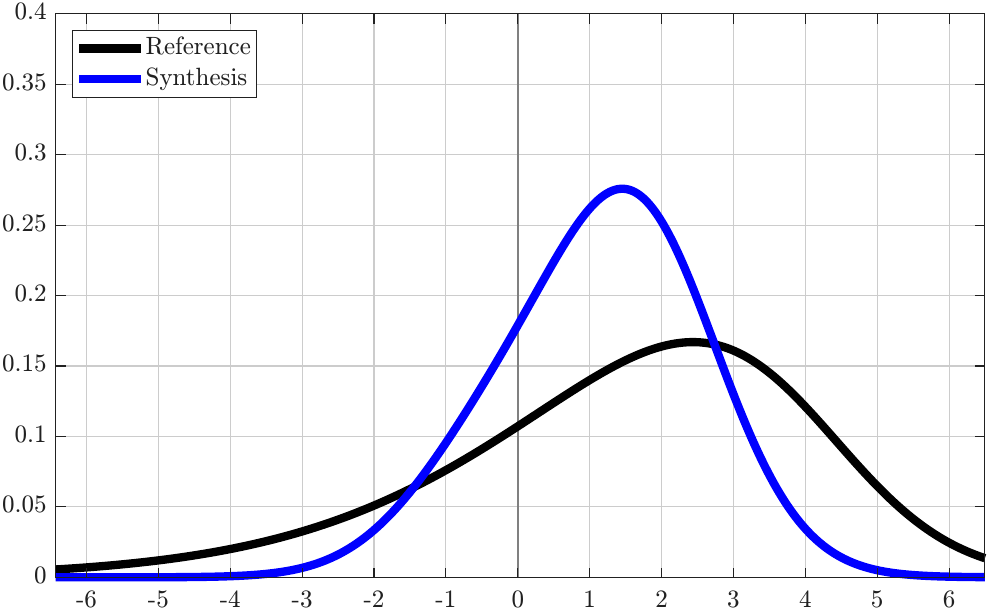} &
\includegraphics[width=.49\textwidth]{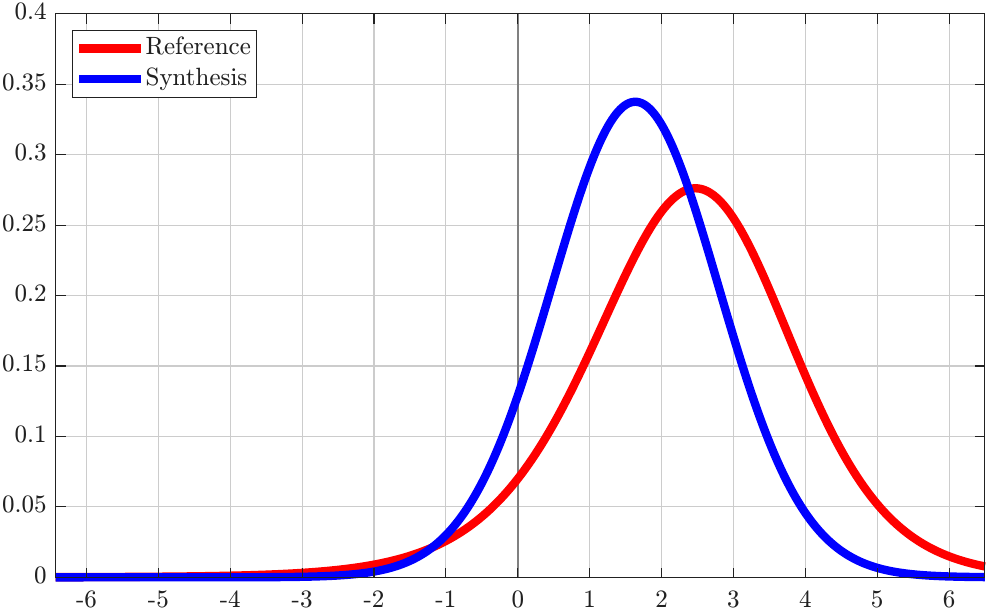}\\
\end{tabular}
\vspace{-3pt} 

\begin{tabular}{p{\textwidth}}\scriptsize
\textsc{Note:} In the left panel, the solid black line denotes the Blackbook Reference p.d.f.; in the right panel, the solid red line denotes the SPF Reference p.d.f. In both panels, the blue line denotes the Scenario Synthesis p.d.f.
\end{tabular}
\end{figure}

Using the Blackbook density as the reference $p(y)$, the Synthesis improves only modestly on the Baseline. The EMR rises from $0.40$ to $0.43$, and the ESS from roughly 58\% to 69\% (left column of Figure~\ref{fig:ScenarioSynthesis2007GDP} and Blackbook panel of Table~\ref{tab:ScenarioSynthesis2007GDP}). The scenario set provides limited coverage of macroeconomic risk: the Blackbook Reference has substantial left-tail risk, while the scenarios remain concentrated around the Baseline.

This shortfall reflects both scenario design and model structure. FRB staff recognized downside risks, but the two downside scenarios were mild relative to the Blackbook density and, ultimately, to the downturn that followed. Moreover, because all scenarios were built with FRB/US, none incorporated the nonlinear amplification mechanisms of a systemic financial crisis. This does not mean that a linear framework such as FRB/US makes severe outcomes impossible: a large enough shock can generate a deep recession. The difficulty is that doing so may force implausible co-movements among inflation, interest rates, and other variables because the model lacks the nonlinear amplification mechanisms---fire sales, collateral spirals, and bank runs---through which a crisis propagates. For such scenarios, a nonlinear model is therefore preferable; see Section~\ref{sec:economics}. 

Next, we use the SPF consensus predictive density as the reference $p(y)$. As shown in the SPF panel of Table~\ref{tab:ScenarioSynthesis2007GDP}, the same Tealbook scenario set receives markedly different weights. The synthesis assigns most weight to the upside scenarios, while the adverse scenarios receive essentially none. By contrast, under the Blackbook Reference, the Credit Crunch scenario receives a weight of approximately $0.24$. This reflects how the two references allocate probability mass: the Blackbook calls for the most severe available downside scenario, while the SPF emphasizes relatively optimistic scenarios. It is worth noting that, regardless of the reference being matched, the ``Greater housing correction'' scenario receives little weight. This does not imply that housing risks were unimportant; rather, the scenario is too close to the Baseline to help match the left tail of the Blackbook Reference.

\begin{table}[htbp]  \caption{Scenario Synthesis --- Summary statistics and weights}\label{tab:ScenarioSynthesis2007GDP} \centering
\footnotesize{2008 Q4/Q4 GDP growth -- Dec. 2007 Tealbook}\\[3pt]
\centering \small \setlength{\tabcolsep}{.0\textwidth} 
\begin{tabular}{C{.05\textwidth}L{.35\textwidth}|C{.1\textwidth}C{.1\textwidth}C{.1\textwidth}|C{.1\textwidth}C{.1\textwidth}C{.1\textwidth}}
\hline\hline
& & \multicolumn{3}{c}{Blackbook Reference} \vline & \multicolumn{3}{c}{SPF Reference} \\ \hline
{$j$} & {Scenario $\scn_j$} & {ESS\%} & {EMR} & {$\alpha_j^*$} & {ESS\%} & {EMR} & {$\alpha_j^*$} \\\hline
0 & Baseline                          & 57.9 & 0.396 & 0.238 & 61.1 & 0.419 & 0.313 \\
1 & Greater housing correction        & 54.0 & 0.383 & 0.009 & 47.4 & 0.379 & 0.007 \\
2 & Credit crunch                     & 37.1 & 0.313 & 0.238 & 14.3 & 0.221 & 0.013 \\
3 & Stronger domestic demand          & 61.5 & 0.407 & 0.238 & 76.7 & 0.456 & 0.313 \\
\grigio 4 & \grigio{With better export performance} & & & & & & \\
5 & More room to grow                 & 62.7 & 0.410 & 0.238 & 83.2 & 0.469 & 0.313 \\
6 & Greater cost pressure             & 56.9 & 0.393 & 0.011 & 57.1 & 0.408 & 0.009 \\
7 & Market-based FFR                  & 60.3 & 0.404 & 0.027 & 71.0 & 0.443 & 0.031 \\\hdashline
  & Synthesis                         & 69.3 & 0.430 &       & 74.5 & 0.452 &       \\\hline
\end{tabular}

\begin{tabular}{p{.99\textwidth}} \scriptsize
\textsc{Note:} ESS\% denotes the effective sample size as a percentage of the total sample. EMR is the Expected Misclassification Rate for each scenario. $\alpha_j^*$ are the optimal synthesis weights, which sum to one across the scenarios included in the synthesis. Scenario 4 is grayed out because it is excluded from the synthesis.
\end{tabular}
\end{table}

The scenario set also fails to span the SPF Reference well: $\demr = 0.5 - 0.452 = 0.048$, or 25.5\% when computed as $100-\mathrm{ESS}$. This residual divergence is smaller than with the Blackbook Reference, where $\demr = 0.070$ (30.7\% as $100-\mathrm{ESS}$), but this should not be interpreted as evidence that the SPF provided a better risk assessment. Concordance is only as informative as the reference being matched: the SPF is easier to span precisely because it assigns little probability to adverse outcomes. The more relevant diagnostic is where the mismatch occurs. In contrast to the Blackbook case, the right column of Figure~\ref{fig:ScenarioSynthesis2007GDP} shows that the incompleteness is concentrated in the right tail, as the available scenarios do not generate sufficiently optimistic outcomes to match the SPF Reference.

\paragraph{Why the Two Syntheses Diverge.}
The divergence arises because the same scenario set is evaluated against references that place probability mass in different regions: the Blackbook assigns much more mass to the left tail, while the SPF assigns little probability to adverse outcomes. The scenario set spans both references imperfectly, but in different regions. The residual divergence is larger against the Blackbook ($0.070$) than against the SPF ($0.048$), but the more informative distinction is the location of the gap: the scenarios are not severe enough for the Blackbook reference and not optimistic enough for the SPF reference. The diagnostic value lies less in the small numerical difference than in the location of the mismatch, which the Synthesis makes transparent by linking reference distributions to weights and fit diagnostics.

This comparison also highlights the value of models that incorporate financial conditions when assessing macroeconomic risks. As discussed in Section~\ref{sec:fed_densities}, Growth-at-Risk approaches formalize the link between financial variables and downside risk. The point is not that a statistical model would necessarily have outperformed expert judgment in real time: the NY Fed's judgmental Blackbook density captured the elevated downside risk that the SPF missed. Rather, a reference informed by financial conditions, as the Blackbook density was, makes the scenario-set gap visible. Against such a reference, the Scenario Synthesis flags systematically and quantitatively that the available scenarios understate left-tail risk.

Overall, this exercise illustrates how the Scenario Synthesis provides a unified framework for evaluating whether a given scenario set adequately captures alternative risk assessments and diagnosing the sources of any mismatch.

\subsection{December 2018: Univariate Analysis}\label{sec:2018_uni}

In this section, we apply the Scenario Synthesis to the alternative scenarios reported in the December 2018 Tealbook \citep{Dec2018Tealbook}. The December 2018 Tealbook provides a useful ``normal-times'' benchmark, as macroeconomic and financial conditions were relatively stable. Two statistical reference densities were available at the time: the Time-Varying Macroeconomic Risk (TVMR) density reported in the Tealbook and the New York Fed's Outlook-at-Risk (OaR) density.\footnote{Although OaR was first published in 2023, historical estimates are available, and the underlying methodology was already well established in 2018, which is why we will use it in this section.} We begin with one-year-ahead GDP growth and then extend the analysis to the joint distribution of GDP growth and core PCE inflation.

\subsubsection{Scenario Set and Modeling Framework}\label{sec:2018_scenarios}
Unlike December~2007, the December~2018 Tealbook lets us compare the same scenario set with both the NY Fed's OaR density and the Board's own TVMR density. It therefore provides a setting in which the narrative and statistical approaches to risk can be evaluated jointly within the Federal Reserve System.

The December 2018 Tealbook features five alternative scenarios (Table \ref{tab:Baseline2018}). The ``Financial-based recession'' scenario ($\scn_1$) examines a situation in which a correction in financial market valuations, combined with constraints on financial intermediaries, triggers a recession. The ``Stronger Supply Side'' scenario ($\scn_2$) assumes more favorable supply conditions than in the Baseline, with a smaller output gap and faster potential growth. In contrast, the ``Supply constraints'' scenario ($\scn_3$) considers the consequences of prolonged labor-market tightness. In this scenario the growth is the same as the baseline, but as we will discuss later, inflation is higher.\footnote{In this univariate analysis, $\scn_3$ is indistinguishable from the Baseline and hence we drop it.  We will be able to attach weight when considering growth and inflation jointly in Section \ref{sec:Dec2018_GDPPCE}. The Tealbook also presents a sixth scenario, ``Lower Oil Prices,'' which we exclude because its implications for 2019 GDP growth are identical to the Baseline and differ only marginally (0.1 percentage point) for core PCE inflation.} The ``Greater Interest Rate Sensitivity'' scenario ($\scn_4$) assumes that household and business spending respond more strongly to interest rate increases than in the Baseline. Finally, the ``Foreign Slowdown'' scenario ($\scn_5$) simulates the effects of weaker foreign activity and an appreciation of the dollar.

\begin{wraptable}{r}{.6\textwidth} \setstretch{1} \vspace{-12pt}
\caption{Baseline and Alternative Scenarios} \label{tab:Baseline2018}  \small \centering 
\footnotesize{Dec. 2018 Tealbook --- One-year-ahead GDP growth projection}\\[2pt]
\begin{tabular}{L{.005\textwidth} L{.025\textwidth} L{.325\textwidth}R{.075\textwidth}R{.075\textwidth}R{.075\textwidth}R{.005\textwidth}} \hline\hline
&{$j$} &   {Scenario $\scn_j$}      &   {P15} &   {P50} &   {P85}      &\\ \hline
& 0     &   Baseline                              &   1.2 &   2.4 &   3.9 &\\
& 1   &   Financial-based recession             &       & $-$0.7  &   \\
& 2   &   Stronger supply side                  &       &   3.1   &   \\
& \grigio 3   &   \grigio{Supply constraints}           &       &   \grigio{2.4}   &   \\
& 4   &   Greater interest rate sensitivity     &       &   1.5   &     &\\
& 5   &   Foreign slowdown                      &       &   1.6  &    &\\ \hline      
\end{tabular}

\begin{tabular}{p{.585\textwidth}} \scriptsize
\textsc{Note:} The baseline projection for 2019 is on page 88, while the alternative scenarios are on page 84. P50 is the point forecast (Baseline and Scenarios), while P15 and P85 are the bounds of the 70\% interval.   
\end{tabular}\vspace{-12pt}
\end{wraptable}
Scenarios $\scn_2$ and $\scn_4$ are constructed using FRB/US (so $M_2 = M_4 = M_0$). Scenario $\scn_1$ uses a Gertler--Karadi-type model \citep{GertlerKaradi2011} with a richer financial sector, $\scn_3$ relies on a calibrated New Keynesian DSGE model with labor-market search frictions similar to \citet{GertlerSalaTrigari2008}, and $\scn_5$ is based on SIGMA, a calibrated multicountry DSGE model maintained at the Federal Reserve Board. The assumptions underlying the scenarios are non-overlapping and correspond to distinct economic mechanisms. 

\subsubsection{Reference and Baseline Distributions}\label{sec:2018_TVMR}

For the December 2018 Tealbook, we consider two alternative reference distributions: the TVMR density reported in the Tealbook and the OaR density produced by the New York Fed.  
\begin{wraptable}{r}{.6\textwidth}\setstretch{1} \vspace{-15pt}
\caption{Skew-$t$ parameters -- Dec. 2018} \label{tab:RB_SKTparam2018}
\centering \small 
\begin{tabular}{L{.01\textwidth}L{.19\textwidth}C{.1\textwidth}C{.1\textwidth}C{.1\textwidth}C{.1\textwidth}}
\hline \hline
&& lc & sc & sk & df \\\hline
&OaR Reference  & 2.5 & 1.3 & -0.3 & 3.0  \\
&TVMR Reference & 2.2 & 1.1 & 0.4 & 30.8 \\  
&Baseline       & 2.4 & 1.3 & 0.0 & 50.0 \\\hline
\end{tabular} 

\begin{tabular}{p{.6\textwidth}} \scriptsize
\textsc{Note:} This Table shows the \skt fitted parameters for the OaR and TVMR References and the Baseline. The parameters reported in the table correspond to the location (lc), scale (sc), skewness (sk), and degrees of freedom (df).
\end{tabular}
\vspace{-12pt}
\end{wraptable}
In practice, both the Tealbook and the New York Fed report predictive percentiles (P5, P15, P50, P85, and P95 for TVMR; P10, P25, P50, P75, and P90 for OaR). To obtain full reference distributions, we fit a \skt density to these percentiles. The Baseline is constructed in the same way as in Section~\ref{sec:2007}.

\begin{figure}[h]\caption{Reference -- Baseline -- Scenarios}\label{fig:RBS_GDP_2018}
\centering
\textsc{\small 2019 Q4/Q4 GDP growth -- Dec. 2018 Tealbook}\\ \smallskip
\begin{tabular}{C{.5\textwidth}C{.5\textwidth}}
\footnotesize \sc OaR Reference & \footnotesize  \sc TVMR Reference \\ 
\includegraphics[width=.49\textwidth]{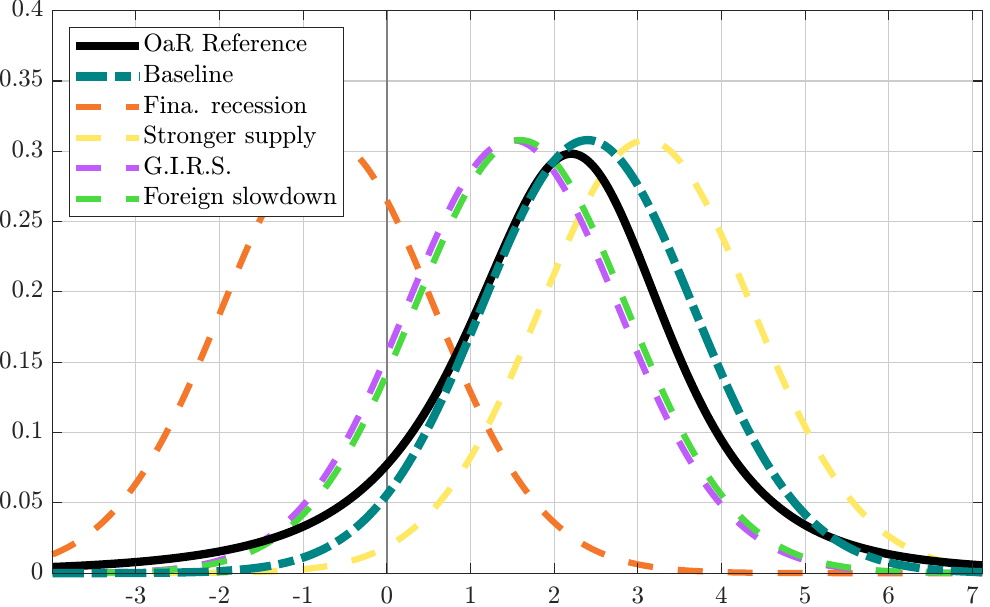} &
\includegraphics[width=.49\textwidth]{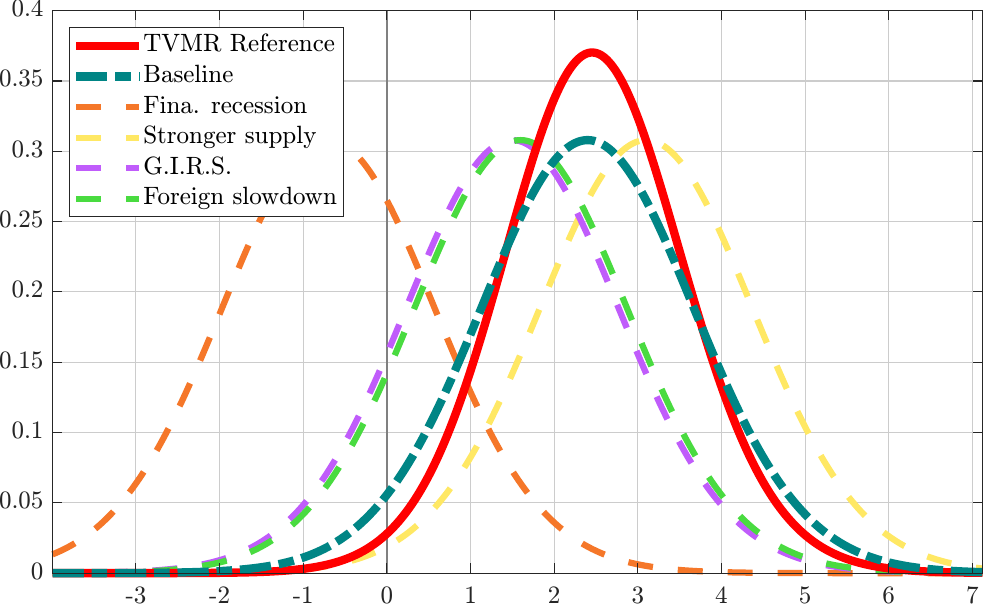}\\
\end{tabular}

\begin{tabular}{p{\textwidth}}\scriptsize
\textsc{Note:} In the left panel, the solid black line denotes the OaR Reference p.d.f. In the right panel, the solid red line denotes the TVMR Reference p.d.f. In both panels, the teal dash-dotted line denotes the Baseline p.d.f., while the dashed lines show the scenario p.d.f.s. 
\end{tabular}\vspace{-1pt} 
\end{figure}

Figure~\ref{fig:RBS_GDP_2018} shows that the Baseline is broadly aligned with both References. However, Table~\ref{tab:RB_SKTparam2018} reveals meaningful differences in shape. Relative to the Baseline, the OaR Reference exhibits fatter tails and modest negative skewness, whereas the TVMR Reference is somewhat less dispersed and displays mild positive skewness. Thus, even in a relatively tranquil period, the two statistical approaches imply different assessments of the risks around the baseline outlook.

\subsubsection{Scenario Synthesis}\label{sec:ScenarioSynthesis2018}

\begin{figure}[h]\caption{Scenario Synthesis --- Probability density functions}\label{fig:ScenarioSynthesis2018GDP}
\centering
\textsc{\small 2019 Q4/Q4 GDP growth -- Dec. 2018 Tealbook}\\ \smallskip
\begin{tabular}{C{.5\textwidth}C{.5\textwidth}}
\footnotesize \sc OaR Reference & \footnotesize  \sc TVMR Reference \\ 
\includegraphics[width=.495\textwidth]{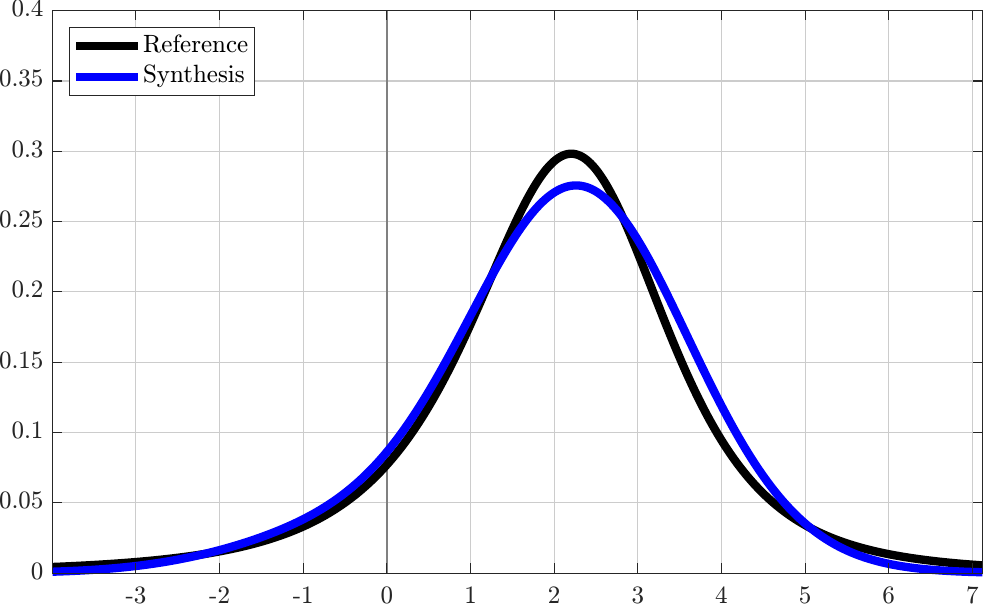} &
\includegraphics[width=.495\textwidth]{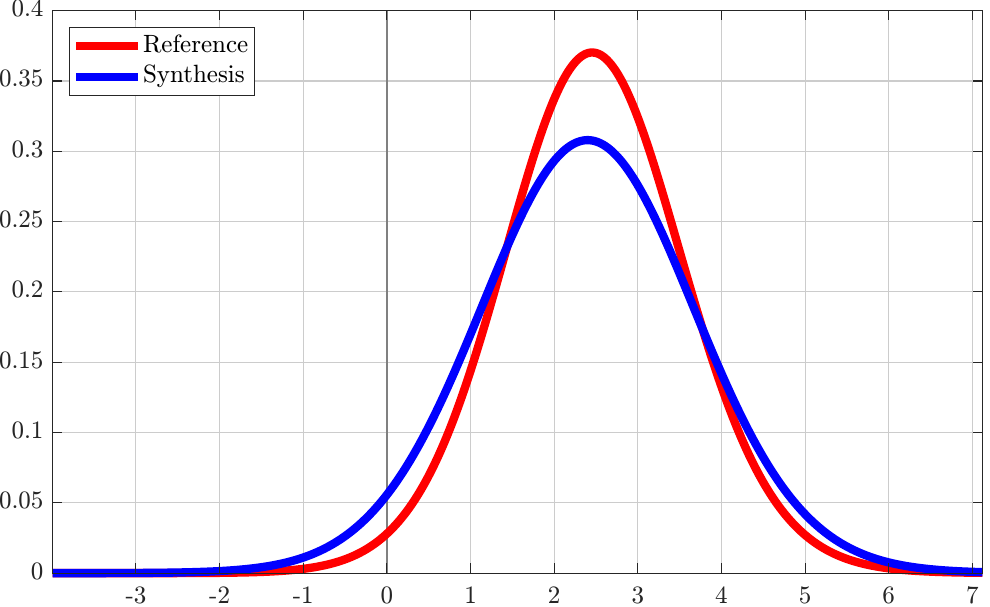}\\
\end{tabular}
\vspace{-3pt} 

\begin{tabular}{p{\textwidth}}\scriptsize
\textsc{Note:} In the left panel, the solid black line denotes the OaR Reference p.d.f.; in the right panel, the solid red line denotes the TVMR Reference p.d.f. In both panels, the blue line denotes the Scenario Synthesis p.d.f.
\end{tabular}
\end{figure}

Figure~\ref{fig:ScenarioSynthesis2018GDP} reports the resulting syntheses. When the Synthesis is constructed to match the OaR Reference, the fit is excellent, with only a minor underestimation of probability mass near the center of the distribution. When the Synthesis is constructed to match the TVMR Reference, the fit is also strong, but the decomposition is much less informative: because the TVMR Reference lies so close to the baseline, the Synthesis places most of the weight on the Baseline and only limited weight on the alternative scenarios.

The summary statistics in Table~\ref{tab:ScenarioSynthesis2018GDP} confirm these visual impressions. In both cases, the scenario set spans the reference distribution well. However, the decomposition implied by the two references differs markedly. Because the TVMR Reference is extremely close to the Baseline, the Baseline receives a weight of 0.87, while all scenarios other than ``Stronger Supply Side'' receive negligible weight. By contrast, when the reference is OaR, whose fitted density exhibits fatter tails, the resulting weights provide a richer characterization of risks around the outlook.

Under the OaR Reference, we can summarize the December 2018 risk assessment as follows: ``Risks to the outlook are modestly skewed to the downside. The Synthesis assigns almost half of its weight (about 48\%) on the Baseline materializing one year ahead, about 15\% weight on a modest upside scenario (for example, due to stronger supply conditions), roughly 30\% on mild downside scenarios, and about 7\% to a financially driven recession.''

\begin{table}[htbp]  \caption{Scenario Synthesis --- Summary statistics and weights}\label{tab:ScenarioSynthesis2018GDP} \centering
\footnotesize{2019 Q4/Q4 GDP growth -- Dec. 2018 Tealbook}\\[3pt]
\centering \small \setlength{\tabcolsep}{.0\textwidth} 
\begin{tabular}{C{.05\textwidth}L{.35\textwidth}|C{.1\textwidth}C{.1\textwidth}C{.1\textwidth}|C{.1\textwidth}C{.1\textwidth}C{.1\textwidth}}
\hline\hline
& & \multicolumn{3}{c}{OaR Reference} \vline & \multicolumn{3}{c}{TVMR Reference} \\ \hline
{$j$} & {Scenario $\scn_j$} & {ESS\%} & {EMR} & {$\alpha_j^*$} & {ESS\%} & {EMR} & {$\alpha_j^*$} \\\hline
0 & Baseline                          & 90.6  &  0.481 &  0.476 & 91.8  &  0.493 &  0.871 \\
1 & Financial-based recession         & 15.5  &  0.229 &  0.069 & ~0.2  &  0.137 &  0.003 \\
2 & Stronger supply side              & 62.7  &  0.434 &  0.153 & 62.9  &  0.464 &  0.086 \\
\grigio 3   &   \grigio{Supply constraints}     & & & & & & \\
4 & Greater interest rate sensitivity & 79.2  &  0.465 &  0.133 & 33.3  &  0.426 &  0.018 \\
5 & Foreign slowdown                  & 83.0  &  0.471 &  0.169 & 39.9  &  0.438 &  0.022 \\\hdashline
  & Synthesis                         & 96.0  &  0.492 &        & 88.6  &  0.491 &        \\\hline
\end{tabular}

\begin{tabular}{p{.99\textwidth}} \scriptsize
\textsc{Note:} ESS\% denotes the effective sample size as a percentage of the total sample. EMR is the Expected Misclassification Rate for each scenario. $\alpha_j^*$ are the optimal synthesis weights, which sum to one across scenarios 0--5. Scenario 3 is grayed out as it is excluded from the synthesis. 
\end{tabular}
\end{table}

Compared to the December 2007 case, the December 2018 scenario set spans the reference distribution well, consistent with a period in which alternative risk assessments were broadly aligned.

\subsection{December 2018: Multivariate Analysis}\label{sec:Dec2018_GDPPCE}

The analysis so far has focused on GDP growth alone. This is sufficient to build intuition, but policymakers care about multiple variables at once. In particular, the Fed's dual mandate makes monetary policy a trade-off between output and inflation risks, requiring a joint assessment. We therefore extend the analysis to a bivariate setting with one-year-ahead GDP growth and core PCE inflation.

This extension has an important implication: all five scenarios become potentially informative. The ``Supply constraints'' scenario, excluded in the univariate case because it coincided with the Baseline GDP median, introduces a distinct inflationary configuration once inflation is included and therefore contributes new information to the synthesis (Table~\ref{tab:Baseline2018multi}).

\begin{table}[ht]\caption{Baseline and Alternative Scenarios} \label{tab:Baseline2018multi}  \small \centering 
\footnotesize{2019 Q4/Q4 GDP growth and core PCE price inflation projections -- Dec. 2018 Tealbook}\\[2pt]

\begin{tabular}{L{.005\textwidth} L{.025\textwidth} L{.325\textwidth}R{.001\textwidth}|*{3}{R{.075\textwidth}}R{.05\textwidth}|*{3}{R{.075\textwidth}}R{.05\textwidth}} \hline\hline
& & & &  \multicolumn{4}{c}{GDP}  \vline & \multicolumn{4}{c}{Core PCE price inflation} \\ \hline
&  {$j$}    &     {Scenario $\scn_j$}        &&{P15}  &  {P50}&  {P85}&& {P15}&  {P50}&  {P85}& \\ \hline
&0&      Baseline                               && 1.2   &  2.4     &   3.9   && 1.2   &  2.0   &  2.8   &  \\
&1&      Financial-based recession              &&       &$-$0.7 &        &&       &   1.9   &     &     \\
&2&      Stronger supply side                   &&       &  3.1     &        &&       &   2.0   &     &     \\
&3&      Supply constraints                     &&       & 2.4     &        &&       &   2.6   &     &     \\
&4&      Greater interest rate sensitivity      &&       & 1.5     &        &&       &   2.0   &     &  \\
&5&      Foreign slowdown                       &&       & 1.6     &        &&       &   1.7   &     &  \\ \hline  
\end{tabular}
\begin{tabular}{p{.906\textwidth}} \scriptsize
\textsc{Note:} The baseline projection for 2019 is on page 88, while the alternative scenarios are on page 84. P50 is the point forecast (Baseline and Scenarios), while P15 and P85 are the bounds of the 70\% interval.    
\end{tabular}
\end{table}

\subsubsection{Reference and Baseline Distributions}

For the multivariate analysis, we focus on the OaR density because Section~\ref{sec:2018_uni} shows that TVMR is extremely close to the Baseline and therefore less informative. We construct joint Reference and Baseline distributions by combining the GDP growth and core PCE marginals with a copula dependence structure, calibrating the correlation matrix from a VAR and imposing no tail dependence (see Appendix~\ref{sec:multivariate_analysis}). The marginals are constructed as in the univariate exercise; for core inflation, however, we had to convert the OaR headline CPI density into a core PCE density as described in \ref{sec:CPItoPCEPx}.

\begin{figure}[h]\caption{Reference, Baseline, and Scenarios -- Dec. 2018 Tealbook}\label{fig:RBS_2018_multi_marginals}
\centering \footnotesize \smallskip
\begin{tabular}{C{.5\textwidth}C{.5\textwidth}}
\sc 2019 Q4/Q4 GDP growth & \sc 2019 Q4/Q4 Core PCE price inflation \\
\begin{tikzpicture}
  \node[anchor=south west,inner sep=0] (image) at (0,0) {\includegraphics[width=.49\textwidth]{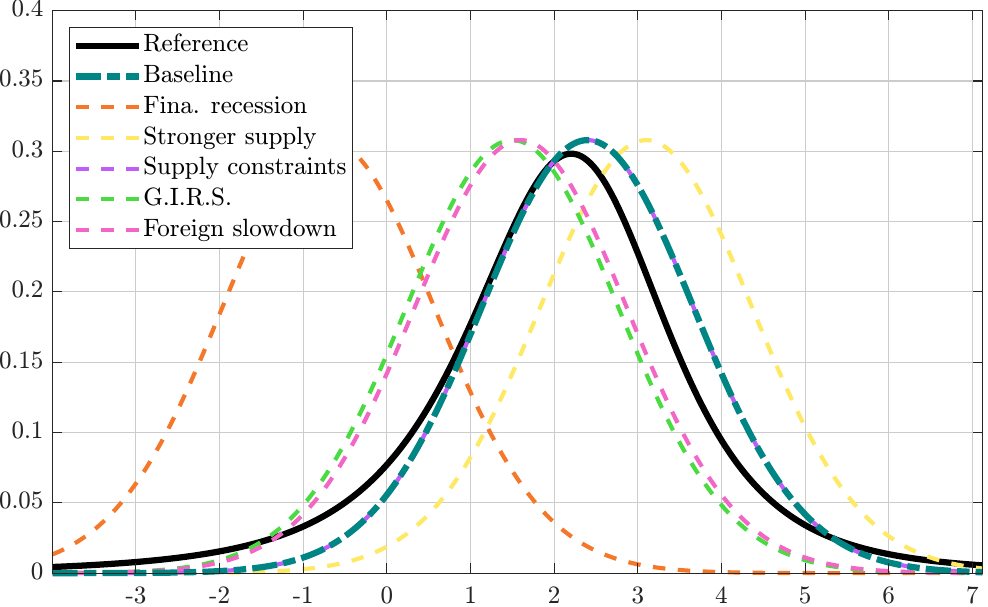}};         
    \begin{scope}[x={(image.south east)},y={(image.north west)}]                
    \node[anchor=south west,inner sep=0,fill=white] (image) at (0.5,0.79) 
            {\setlength{\tabcolsep}{.0\textwidth} \tiny
      \begin{tabular}{|C{.005\textwidth}L{.07\textwidth}R{.035\textwidth}R{.035\textwidth}R{.035\textwidth}R{.035\textwidth}R{.005\textwidth}|}\hline
      &            & lc  &  sc &  sk     &   df  &\\\hline                
      &Reference   & 2.6 & 1.3 & -0.3 & 3.0 &\\
      &Baseline    & 2.4 & 1.3 & 0.0  & 50.0  &\\\hline    
      \end{tabular}};  
  \end{scope}
\end{tikzpicture} &

\begin{tikzpicture}
  \node[anchor=south west,inner sep=0] (image) at (0,0) {\includegraphics[width=.49\textwidth]{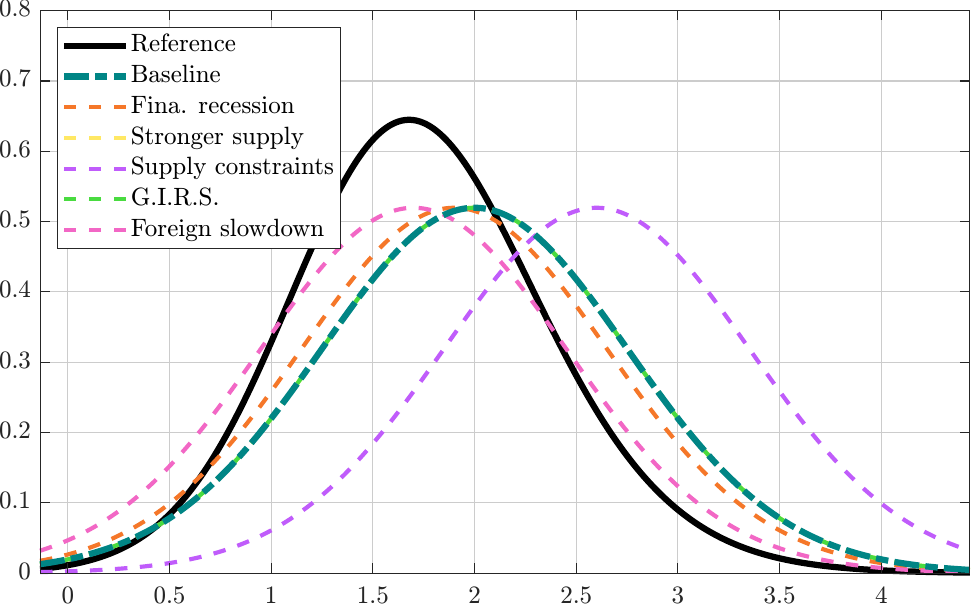}};
    \begin{scope}[x={(image.south east)},y={(image.north west)}]                
    \node[anchor=south west,inner sep=0,fill=white] (image) at (0.5,0.79) 
            {\setlength{\tabcolsep}{.0\textwidth} \tiny
      \begin{tabular}{|C{.005\textwidth}L{.07\textwidth}R{.035\textwidth}R{.035\textwidth}R{.035\textwidth}R{.035\textwidth}R{.005\textwidth}|}\hline
      &            & lc  &  sc &  sk     &   df  &\\\hline                
      &Reference   & 0.9 & 1.4 & 1.0  & 30.0 &\\
      &Baseline    & 2.0 & 0.8 & 0.0  & 50.0  &\\\hline         
      \end{tabular}};  
  \end{scope}
\end{tikzpicture}\\
\end{tabular}

\begin{tabular}{p{\textwidth}}\scriptsize
\textsc{Note:} In both panels, the solid black line denotes the OaR Reference p.d.f. and the teal dash-dotted line denotes the Baseline p.d.f. Dashed lines show the scenario p.d.f.s for GDP growth (left panel) and core PCE price inflation (right panel). The skew-$t$ parameters reported in the inset boxes (upper right of each panel) correspond to the location (lc), scale (sc), skewness (sk), and degrees of freedom (df).
\end{tabular}\vspace{-1pt} 
\end{figure}

Figure~\ref{fig:RBS_2018_multi_marginals} shows the marginals. For GDP growth, the comparison is unchanged from the univariate case: the Reference and Baseline are centered close together, differing mainly in asymmetry and tails. For core inflation, the contrast is much stronger: the Reference is shifted left, much more dispersed, and strongly positively skewed, while the Baseline remains centered near 2 percent and symmetric. The inflation marginal therefore reflects disagreement about both the center and the balance of risks, which affects which scenarios help match the joint Reference.

\begin{figure}[h]\caption{Reference, Baseline, and Scenarios -- Dec. 2018 Tealbook}\label{fig:RBS_2018_multi_bivariate}
\centering \footnotesize \smallskip
\begin{tabular}{C{.5\textwidth}C{.5\textwidth}}
\sc Baseline and Scenarios as point forecasts & \sc Baseline as bivariate distribution \\
\includegraphics[width=.49\textwidth,trim=0cm 0cm 0cm 0cm,clip]{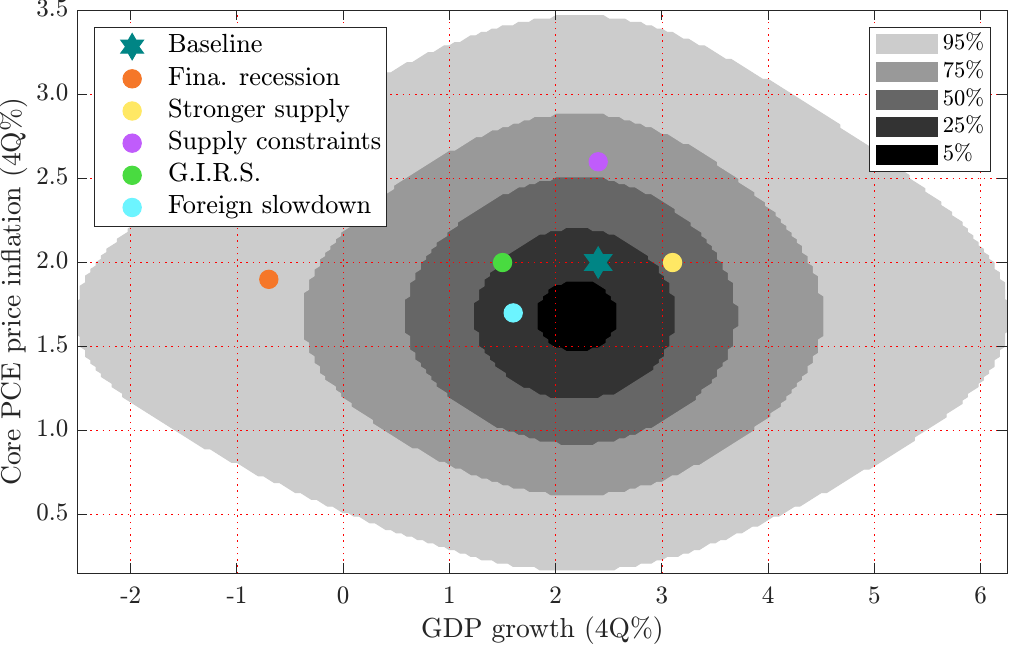}&
\includegraphics[width=.49\textwidth,trim=0cm 0cm 0cm 0cm,clip]{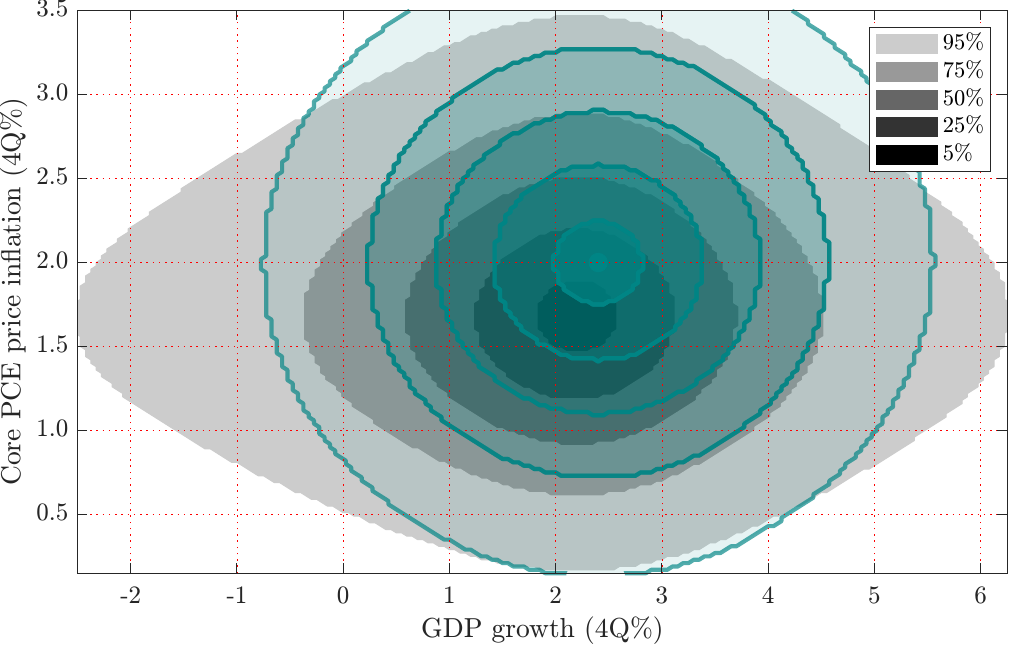}\\
\end{tabular}

\begin{tabular}{p{\textwidth}}\scriptsize
\textsc{Note:} Both panels show joint probability density for GDP growth ($x$-axis) and core PCE inflation ($y$-axis). Contour lines show regions of equal probability density. Color intensity indicates probability concentration. The left panel displays the OaR Reference \pdf with colored points showing the Baseline and alternative scenario forecasts as (GDP, inflation) pairs. Contour regions denote the 5\%, 25\%, 50\%, 75\%, and 95\% highest density regions. The right panel shows only the Baseline joint \pdf (teal contours) overlaid on the Reference \pdf (gray shaded regions).
\end{tabular}\vspace{-1pt} 
\end{figure}

Figure~\ref{fig:RBS_2018_multi_bivariate} adds the bivariate view. The left panel places the Baseline and scenario point forecasts on the Reference contour map: the scenarios are dispersed along GDP but compressed along inflation, so only a subset is well positioned to match the joint shape. The right panel overlays the Baseline joint density, which captures the broad center but not the full geometry, especially along inflation.\footnote{Scenario marginals are constructed as in the univariate exercise and combined using the same correlation structure as the Baseline and Reference.}

\subsubsection{Scenario Synthesis}

\begin{figure}[htbp]
\centering

\begin{minipage}[t]{0.45\textwidth} \centering
  \caption{Scenario Synthesis}\label{fig:ScenarioSynthesis2018GDPmulti} 
  \footnotesize{\sc Probability density function}\\
  \scriptsize{\sc GDP growth and core inflation}\\
  \tiny{\sc Dec. 2018 Tealbook}\\[3pt]
    \includegraphics[width=\textwidth,trim= .5cm .5cm 0cm 0cm, clip]{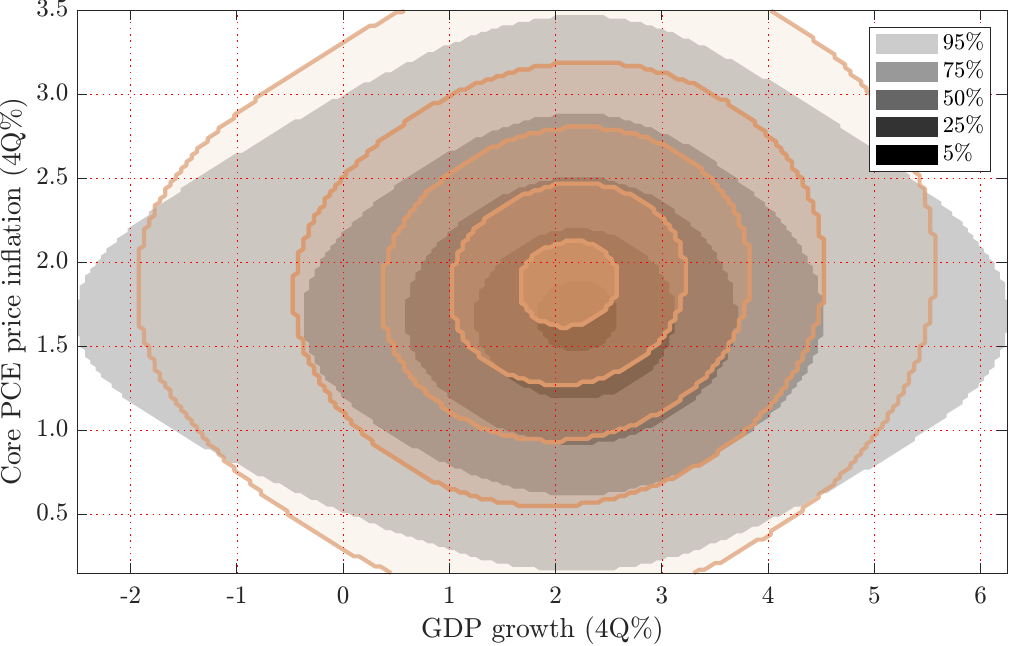} 

\begin{tabular}{p{.975\textwidth}} \scriptsize
\textsc{Note:} This figure shows probability density contours for the joint \pdf of GDP growth ($x$-axis) and core PCE price inflation ($y$-axis) under the Scenario Synthesis distribution (orange contours), overlaid on the joint Reference \pdf (gray shaded regions). Contour regions denote the 5\%, 25\%, 50\%, 75\%, and 95\% highest density regions. 
\end{tabular}

\end{minipage}
\hfill
\begin{minipage}[t]{0.54\textwidth} \centering
  \captionof{table}{Scenario Synthesis}\label{tab:ScenarioSynthesis2018GDPmulti} \centering
  \footnotesize{\sc Summary statistics and weights} \\
  \scriptsize{\sc GDP growth and core inflation}\\
  \tiny{\sc Dec. 2018 Tealbook}\\[3pt]
  \footnotesize 
  \begin{tabular}{C{.05\textwidth}L{.59\textwidth}C{.12\textwidth}C{.12\textwidth}C{.12\textwidth}} \hline\hline
  {$j$} & {Scenario $\scn_j$}           & {ESS\%}  & {EMR}  & $\alpha_j^*$ \\\hline
0   & Baseline                          & 69.5  &  0.459 &  0.376   \\       
1   & Financial-based recession         & 13.1  &  0.225 &  0.060   \\       
2   & Stronger supply side              & 48.2  &  0.415 &  0.152   \\       
3   & Supply constraints                & 17.3  &  0.347 &  0.006   \\       
4   & Greater interest rate sensitivity & 60.8  &  0.444 &  0.030    \\       
5   & Foreign slowdown                  & 70.1  &  0.462 &  0.376   \\\hdashline   
    & Synthesis                         & 79.9  &  0.477 &          \\\hline 
\end{tabular}

\begin{tabular}{p{.975\textwidth}} \scriptsize
\textsc{Note:} ESS\% denotes the effective sample size as a percentage of the total sample. EMR is the Expected Misclassification Rate for each scenario. $\alpha_j^*$ are the optimal synthesis weights, which sum to one across scenarios 0--5.  
\end{tabular}

\end{minipage}

 \end{figure}

Figure~\ref{fig:ScenarioSynthesis2018GDPmulti} and Table~\ref{tab:ScenarioSynthesis2018GDPmulti} report the resulting Scenario Synthesis. The contour plot shows that the Synthesis improves visibly upon the Baseline and tracks the overall shape of the Reference reasonably well, although the fit is far from exact. This improvement is evident both graphically---by comparing Figure~\ref{fig:ScenarioSynthesis2018GDPmulti} with the right panel of Figure~\ref{fig:RBS_2018_multi_bivariate}---and in the summary statistics: the EMR rises from 0.459 for the Baseline to 0.477 for the Synthesis, while the ESS increases from 69.5\% to 79.9\%.

The estimated weights show that this improvement is achieved through a selective reweighting of the scenario set rather than through broad use of all scenarios. The Baseline and the ``Foreign slowdown'' scenario receive equal weight, 0.376 each, and together account for more than three quarters of the synthesis. The next most important component is ``Stronger supply side,'' with weight 0.152, followed by ``Financial-based recession,'' with weight 0.060. By contrast, ``Greater interest rate sensitivity'' receives only 0.03, and ``Supply constraints'' is essentially irrelevant, with weight 0.006.

This pattern is intuitive given the joint Reference. Once output and inflation risks matter jointly, the most useful scenarios match not only the GDP distribution but also the inflation asymmetry. The ``Foreign slowdown'' scenario does this well: weaker growth with lower inflation reproduces the left-shifted inflation distribution without moving GDP far from the Reference mass. By contrast, ``Supply constraints,'' which mainly raises inflation, becomes much less useful. Adding inflation thus makes the scenario set more informative but exact coverage more demanding.

%
%
\section{Practical Considerations}\label{sec:learnings}

The empirical exercises highlight that the Scenario Synthesis is not just a tool for ex post evaluation. It also raises a set of practical questions for real-time policy work: Where do scenarios come from? What should be done when scenarios are reported only as point forecasts or narratives rather than as full predictive distributions? And how should institutions respond when the available scenario set is unable to span the relevant risks? This section discusses these issues.

\subsection{Where do scenarios come from?}

A practical question is how the scenario set is selected. As discussed in Section~\ref{sec:fed_scenarios}, scenarios change from round to round: new risks emerge, old concerns fade, and the selection criteria are usually implicit. Staff ask which risks are salient, which mechanisms the baseline may miss, and which questions policymakers are raising, but these criteria are rarely formalized.

This matters because the scenario set changes from round to round: one cannot track the weight on a named scenario if it may not exist next round. But the Synthesis also reduces the need to communicate risk shifts through scenario turnover. Without it, declining optimism is conveyed by adding downside scenarios or removing upside ones; with it, the same shift appears in the weights on scenarios already on the table. What matters is not only which scenarios are present, but how weight is distributed across them. Turnover remains relevant but less central, and the Synthesis can inform selection by diagnosing missing and redundant scenarios.

A related question is reverse stress testing, which derives the scenarios that would produce a given outcome from exposures, balance sheets, or realized losses. Scenario design is often iterative in this spirit: construct, assess, adjust. A reverse-engineered scenario can be treated as another conditional density $p_j(y)$ once finalized, and the diagnose-and-re-synthesize loop of Section~\ref{sec:2007_incompleteness} follows the same logic. A full treatment is beyond our scope, but the framework is compatible with it.

\subsection{Scenario predictive distributions}\label{sec:ScenarioDistributions}

In our theoretical framework, each scenario provides a full predictive distribution $p_j(y)$, constructed using the appropriate model $M_j$ and assumptions $A_j$. This is the first-best case, because the synthesis can then be implemented directly using the scenario densities.

In practice, scenarios are often presented only as point forecasts or narratives. In this second-best case, the framework can still be implemented by constructing scenario-consistent densities via entropic tilting or related methods that embed the scenario path in the baseline's uncertainty---a feasible workaround, used in Section~\ref{sec:empirics}, but not ideal. The limitation bites when the scenario requires $M_j \neq M_0$: the reason for a different model is usually that the baseline cannot represent the nonlinear mechanism, so a scenario from a nonlinear accelerator or kinked Phillips curve implies asymmetric uncertainty or fat tails that tilting the baseline density may miss. For internal policy analysis, institutions should produce full predictive distributions for scenarios whenever possible, even if external communication emphasizes central paths.

\subsection{Joint risk assessment}\label{sec:joint_risk_assessment}

All the predictive distributions considered so far are marginal densities rather than full joint densities. In Section~\ref{sec:Dec2018_GDPPCE}, we constructed a joint density using a normal copula, that is, by assuming no tail dependence and calibrating the linear correlation across variables. This is a useful workaround, but at best a second-best solution. Ideally, the reference would provide a full joint risk assessment rather than marginals stitched together through auxiliary assumptions; but the literature has progressed far more on marginal tail risks than on their joint distribution. Multivariate applications thus remain feasible and informative but require dependence assumptions beyond the original reference densities.

The bivariate exercise of Section~\ref{sec:Dec2018_GDPPCE} extends directly to more variables and multiple horizons in principle. In practice, however, the dependence assumptions become more demanding in higher dimensions. The statistical reference densities we use are also predominantly one-year-ahead, which limits the horizons we can study. Finally, beyond the bivariate case, the results become harder to display, since the contour plots used above are no longer available. These are constraints of data and communication, not of principle.

\subsection{What to do if the scenario set is incomplete?}\label{sec:2007_incompleteness}

As shown in the December 2007 case study in Section~\ref{sec:2007}, the scenario set was unable to recover the Reference distribution. In that exercise, we treated the scenario set as fixed. In practice, however, the scenario set is not fixed, and the relevant question becomes what to do when it is incomplete.

A natural starting point is statistical. Scenario-set incompleteness is the analogue of ``model-set incompleteness'' in Bayesian econometrics. Bayesian Predictive Synthesis (BPS)---and its decision-guided extension BPDS \citep{TallmanWest2023,ChernisTallmanKoopWest2024}---augments the initial densities with a \bem{\safehavennospace}: a deliberately over-dispersed distribution centered on the mixture, designed to capture outcomes outside the original support, often implemented as an over-dispersed average of the existing densities. The same idea can be applied to scenario analysis by constructing a \safehaven scenario. One concrete formulation is:
\vspace{-12pt}\[
\text{P50}_{\safehaven} = \operatorname{median}_{j=1,\ldots,J} \text{P50}_j, \qquad
\text{P15}_{\safehaven} = \min_{j} \text{P15}_j, \qquad
\text{P85}_{\safehaven} = \max_{j} \text{P85}_j. \vspace{-12pt}
\]
Such a scenario is grounded in BPS theory and guarantees improved spanning, but lacks a clear economic interpretation, limiting its communicative value; \citet{AGLW2025} provide a concrete implementation of this backstop construction to which we refer the reader for details.

An alternative, and often more economically meaningful, approach is to diagnose which risks are missing and design new scenarios---or recalibrate existing ones---to span those regions of the predictive distribution. In practice, the framework is most useful iteratively: start with the existing scenario set, compute the weights and $\demr$, diagnose the gaps, refine the scenarios, and re-synthesize. The loop can also run in reverse: under-covered regions of the reference can guide the conditioning assumptions for new scenarios. Thus, the Synthesis serves not only as an evaluation tool, assessing an existing scenario set, but also as a design tool, guiding improved scenarios.

%
%

\section{Alternative Economic Models}\label{sec:economics}

Baseline models---FRB/US, NAWM~II, and COMPASS---are local approximations calibrated for normal times. They linearize around a steady state and represent financial frictions only partially. In calm periods this is often adequate; in stress episodes it can become misleading. The nonlinear amplification mechanisms that define crises---fire sales, collateral spirals, and wage-price spirals---are precisely the mechanisms that scenario analysis is designed to illuminate, and they require models the baseline excludes.

This is the operational meaning of $M_j \neq M_0$ in Section~\ref{sec:toolkit}. In December~2007, the ``Credit Crunch'' scenario required endogenous credit feedback that FRB/US could only partially capture. In December~2018, the ``Financial-based recession'' scenario used a Gertler--Karadi-type model \citep{GertlerKaradi2011}, while the ``Supply constraints'' scenario had different inflation implications depending on whether the Phillips curve was flat or steep. We discuss these two mechanisms briefly and then introduce the Macroeconomic Model Database as a tool for understanding model uncertainty.

\paragraph{The nonlinear financial accelerator.}
Standard baseline models include a simplified financial sector: credit spreads respond linearly, while the feedback loops that amplify distress---falling collateral tightening credit, contracting output, and further depressing collateral---are absent. \citet{bernanke1999} embed these loops structurally; \citet{kiyotaki1997} formalize a related mechanism through land prices and collateral constraints; and \citet{GertlerKaradi2011} and \citet{gertler2010} extend the accelerator to constrained intermediaries, providing the structural basis for asset purchases and macroprudential regulation \citep{adrian2010}. The implication is direct: when financial conditions are tight and the accelerator is operative, the reference distribution should place more mass in the left tail, and adverse scenarios should receive higher weights---the pattern captured by the Blackbook density in December~2007 and missed by the SPF.

\paragraph{The nonlinear Phillips curve.}
Standard models embed a linear Calvo Phillips curve: inflation responds proportionally to slack regardless of state. The early-2020s inflation surge challenged this assumption, and evidence for a state-dependent slope has grown. In menu-cost models, large shocks raise adjustment frequency and steepen the curve \citep{costain2019, alvarez2023}; \citet{benigno2023} propose an ``Inverse-L'' curve, flat below a labor-market threshold and steep above; and \citet{harding2023} reach a complementary conclusion through quasi-kinked demand. For scenario design, this changes the stakes of supply scenarios. The December~2018 ``Supply constraints'' scenario examined prolonged tightness: under a linear curve, the inflation implications are moderate; under a convex curve, they are sharply higher, calling for a different model. Assigning probability to the scenario therefore requires taking a stand on which portion of the curve the economy occupies---squarely an $M_j \neq M_0$ disagreement (Section~\ref{sec:toolkit}). The 2021--22 run-up illustrates how a mechanism once confined to a scenario can become central to the baseline.

\paragraph{The Macroeconomic Model Database.}
The mechanisms above are well-understood departures from the baseline. In practice, however, the space of plausible models is much larger, differing in price and wage rigidity, expectations, openness, and policy rules. For a given shock, these differences can generate forecast divergences that rival those between the baseline and the most extreme scenario. The Macroeconomic Model Database \citep[MMB;][]{wieland2012,wieland2016} helps navigate this space: it is an open-source archive of more than 150 structural models from academia and policy institutions, re-coded into a common format so impulse responses can be compared like-for-like. For Scenario Synthesis, it makes the $M_j$ dimension concrete: staff can identify which models predict meaningfully different outcomes for a given shock, and therefore which disagreements merit distinct scenarios. The MMB thus provides a principled way to populate $\{M_j\}$ and turn implicit model uncertainty into explicit weights.

%
%

\section{Policy Considerations: Consensus Under Deep Uncertainty}\label{sec:policy}
Policymakers often disagree about both the near-term outlook and long-run concepts such as the natural rate of interest, the equilibrium unemployment rate, or the slope of the Phillips curve. In the U.S., dispersion in the FOMC's Summary of Economic Projections makes this disagreement visible. Similar disagreements persist at other central banks such as within the ECB's Governing Council and the Bank of England's MPC despite common staff briefings, forecasts, and risk assessments.

Such disagreement is natural under \emph{deep uncertainty}---a term used in the operations research and management science literature to describe situations in which decision-makers cannot agree on the model of the system, the relevant outcomes, or the probability distributions over uncertain inputs \citep{WalkerLempertKwakkel2013}. Central banks face deep uncertainty: there is no agreement on how best to represent the economy, transmission mechanisms shift over time, and the distributions governing future outcomes cannot be pinned down with confidence. 

Disagreement under deep uncertainty has three sources:
\begin{enumerate}[label=\emph{\roman*}),leftmargin=2\parindent,itemsep=0pt,parsep=0pt,topsep=0pt]
\item \emph{Shock uncertainty} concerns the assumptions $A_j$. Policymakers may agree on the model but disagree about which shocks have hit the economy, how to interpret recent developments, or which shocks are likely ahead.

\item \emph{Transmission uncertainty} concerns the model component $M_j$. Policymakers may disagree about how the economy works and which mechanisms are operative. As discussed in Section~\ref{sec:economics}, the same shock path can generate different outcomes depending on whether financial acceleration, nonlinear price adjustment, or another form of state dependence is relevant.

\item \emph{Ranking uncertainty} concerns the loss function. Policymakers may agree on shocks and transmission but still weigh outcomes differently: one may worry more about inflation, another about unemployment. Point forecasts obscure this distinction by combining beliefs about the economy with the weighting of risks; predictive densities help separate what is believed from what is prioritized.
\end{enumerate}

Once these sources of disagreement are explicit, the question becomes how to aggregate them into a consensus risk assessment. Scenario Synthesis offers a natural device. Each committee member's view can be represented as a conditional predictive density $p(y \mid A_j,M_j)$, just as scenarios are. Under this view, disagreement is structured information about which models and assumptions members consider most relevant, not noise to be averaged away. 

Rather than reporting point forecasts that combine beliefs with risk rankings, members could work with staff to design scenarios reflecting their preferred model and assumptions. Scenario Synthesis would then aggregate these views into a committee-level risk assessment. In our applications, we use a relatively flat prior and let the reference density discipline the weights through the concordance criterion. More generally, members could bring priors over scenarios, reflecting their subjective probabilities that a given $(M_j,A_j)$ combination is the relevant approximation.

As views and the reference density evolve, the synthesis can be recomputed, providing a transparent record of how the committee's collective assessment changes over time. The residual divergence $\widetilde{\pi}_{pf}$ is especially useful: a large residual indicates that the scenario set does not span the range of committee views and that new scenarios are needed. The framework thus serves not only as an evaluation tool but also as a discipline for scenario design under deep uncertainty.

Finally, it is important to note that Scenario Synthesis is a tool for risk assessment, not a decision rule. It does not prescribe a policy action or close the feedback loop from risk assessment to policy response and realized outcomes: easing in response to downside risk may itself reduce that risk. Embedding scenario analysis in a decision framework that closes this loop is the subject of a complementary literature. \citet{Cairoetal2025} and \citet{Gargaetal2025}, for example, develop risk-adjusted optimal-control policies in which policymakers weigh scenario probabilities and update them as shocks arrive. A key lesson is that the policy path is not a probability-weighted average of scenario-specific policy paths, because beliefs, policy, and outcomes interact.\footnote{Relatedly, the scenarios entering the synthesis are conditional on an assumed policy reaction function, so the weights inherit that conditioning. Pure monetary-policy scenarios---those that vary only the assumed rule---should therefore be interpreted accordingly.}

%
%

\section{Conclusions}\label{sec:conclusions}

Central banks have developed two approaches to monitoring macroeconomic risk---scenario analysis and predictive distributions---that evolved in parallel yet separately. Scenarios provide narratives without probabilities; predictive distributions provide probabilities without narrative. We argue that the two are complements, not substitutes. We provide a historical account of their emergence and introduce the Scenario Synthesis, a framework that connects them by assigning scenario weights consistent with a reference predictive distribution.

The Scenario Synthesis evaluates existing scenario sets and helps improve them. Scenarios are necessarily subjective and lack continuity across rounds; the Synthesis adds quantitative discipline by comparing the weighted combination of the scenarios with the reference distribution. This makes it possible to assess whether scenarios span the relevant risks, identify what is missing, and guide the construction of new scenarios.

Beyond scenario design, the framework offers a structured way to represent disagreement within policy committees. Members may differ not only in how they weight outcomes, but also in their views of shocks and transmission. This helps explain why disagreement persists even under shared information, and why scenario weights can serve as a disciplined summary of heterogeneous views.

{\setstretch{1.15} \small
\bibliographystyle{chicago}
\bibliography{ScenarioSynthesis2025.bib}
\par}
\normalsize

\clearpage

\pagestyle{fancy}
\fancyhf{}
\lhead{\sc Supplementary material\\[-10pt]}
\rhead{\sc Risks and Uncertainty in Monetary Policy\\[-10pt]}
\cfoot{\sc Page \color{Rosso}\thepage\ \color{black}of \pageref{LastPage}}
 \renewcommand{\footrulewidth}{0pt}
 \renewcommand{\headrulewidth}{0pt}

\gdef\thesection{Appendix \Alph{section}}
\gdef\thesubsection{\Alph{section}.\arabic{subsection}}
\gdef\thefigure{\Alph{section}\arabic{figure}}
\gdef\theequation{\Alph{section}\arabic{equation}}
\gdef\thetable{\Alph{section}\arabic{table}}
\setcounter{page}{1}
\setcounter{table}{0}%
\setcounter{figure}{0}%
\setcounter{equation}{0}%
\setcounter{section}{0}%

\setstretch{1.25}

\makeatletter
\renewcommand\normalsize{%
   \@setfontsize\normalsize{11}{13.6}%
   \abovedisplayskip 8\p@ \@plus4\p@ \@minus4\p@
   \abovedisplayshortskip \z@ \@plus3\p@
   \belowdisplayshortskip 5\p@ \@plus3\p@ \@minus3\p@
   \belowdisplayskip \abovedisplayskip
   \let\@listi\@listI}
\renewcommand\small{%
   \@setfontsize\small{9.5}{11.5}%
   \abovedisplayskip 6\p@ \@plus3\p@ \@minus3\p@
   \abovedisplayshortskip \z@ \@plus2\p@
   \belowdisplayshortskip 4\p@ \@plus2\p@ \@minus2\p@
   \belowdisplayskip \abovedisplayskip}
\renewcommand\footnotesize{%
   \@setfontsize\footnotesize{9}{11}%
   \abovedisplayskip 5\p@ \@plus2\p@ \@minus2\p@
   \abovedisplayshortskip \z@ \@plus\p@
   \belowdisplayshortskip 3\p@ \@plus\p@ \@minus2\p@
   \belowdisplayskip \abovedisplayskip}
\renewcommand\scriptsize{\@setfontsize\scriptsize{7.5}{9.5}}
\renewcommand\tiny{\@setfontsize\tiny{6}{7}}
\renewcommand\large{\@setfontsize\large{12}{14}}
\renewcommand\Large{\@setfontsize\Large{14}{18}}
\renewcommand\LARGE{\@setfontsize\LARGE{17}{22}}
\renewcommand\huge{\@setfontsize\huge{20}{25}}
\renewcommand\Huge{\@setfontsize\Huge{24}{30}}
\makeatother
\normalsize

\begin{center} \color{Rosso}
\rule{\textwidth}{1.5pt}\\
\textit{Supplementary material for the paper:} \\

\textbf{\Huge Risks and Uncertainty\\[6pt] in Monetary Policy}\\
\rule{\textwidth}{1pt}\\ [18pt] 

\color{black}
\begin{tabular}{cp{2cm}c}
\normalsize \sc Tobias Adrian &&  \normalsize \sc Domenico Giannone \\[-5pt]
\small International Monetary Fund &&\small Johns Hopkins University\\[-5.5pt]
\footnotesize tadrian@imf.org &&\footnotesize domenico.giannone@jhu.edu\\[9pt]
\normalsize \sc Matteo Luciani &&  \normalsize \sc Mike West\\[-5pt]
\small Federal Reserve Board &&\small Duke University\\[-5.5pt]
\footnotesize matteo.luciani@frb.gov &&\footnotesize mike.west@duke.edu\\[9pt]
\end{tabular}

\end{center}

\renewcommand{\thefootnote}{ } 

\footnotetext{\noindent \textsc{Disclaimer:} The views expressed in this paper are those of the authors and do not necessarily reflect the views and policies of the Board of Governors, the Federal Reserve System, or the International Monetary Fund, its Management, or its Executive Directors.} 

\gdef\thefootnote{(\roman{footnote})}

%
%

\section{Scenario synthesis implementation}\label{sec:scenario_synthesis_implementation}
This appendix summarizes the main theoretical and computational steps behind the empirical implementation of the Scenario Synthesis. We first describe the univariate implementation and then turn to the multivariate case used in the joint analysis of GDP growth and core PCE inflation.

\subsection{Univariate analysis}
In the univariate applications, the synthesis is implemented as follows:
\begin{enumerate}[label=S\arabic*:,itemsep=0pt,parsep=0pt,topsep=0pt]
   \item Generate a large random sample $\y^i$, $(i=\seq 1n)$, from the reference $p(\y)$, to define an importance sample for Monte Carlo evaluation of the baseline $p_0(\y)$ and the scenarios $p_j(\y).$ 
   \item Evaluate baseline  IS weights $w_0^i \propto p_0(\y^i)/p(\y^i)$, subject to normalization. 
   \item Evaluate the scenario p.d.f.s $p_j(\y)$ for $j>0$ as in Step 2 now applied to scenario p.d.f.s $p_j(\y)$ instead of the baseline $p_0(\y).$ For each $\scnj$ this delivers normalized IS weights $ w_j^i $ on the reference sample values.  
   \item Compute synthesis weights by maximising 
   \[\{\alpha_j^*\}_{j=0}^J
  =
  \argmax_{\substack{\alpha_j>0,\ \alpha_0 \ge \alpha_j \\
  \sum_{j=0}^J \alpha_j = 1}}
  \left[
  \log\{\emr(\balpha)\}
  +
  \epsilon\sum_{j=0}^J \log(\alpha_j)
  \right]\]
  where
 \[ \emr(\balpha) =  \frac{1}{n}\sum_{i=1}^n \frac{w_f^i(\balpha)}{w_f^i(\balpha) + w_p^i}, \]
$w_f^i(\balpha) = \sum_{j=0}^J \alpha_j w_j^i$ are the synthesis IS weights and $w_p^i = 1/n$ are the (uniform) reference weights.        
\end{enumerate}

\subsection{Multivariate analysis}\label{sec:multivariate_analysis}
\begin{enumerate}[label=\~S\arabic*:,itemsep=0pt,parsep=0pt,topsep=0pt]
   \item Generate a large random sample $\y^i$, $(i=\seq 1n)$, from the Reference distribution using the skew-$t$ copula approach, as described in Appendix~\ref{sec::skew_t_copula_sampling}.
   \item Evaluate baseline IS weights:
      \[
         w_0^i \propto \frac{p_0(\y^i)}{p(\y^i)},
      \]
      subject to normalization, where $p_0(\y)$ and $p(\y)$ are computed using the exact parametric copula density formula in Appendix~\ref{sec:copula_density_evaluation}.
   \item Evaluate scenario IS weights:
      \[
         w_j^i \propto \frac{p_j(\y^i)}{p(\y^i)},
      \]
      subject to normalization for each scenario $j$, using the exact parametric copula density.
   \item Compute synthesis weights as in S4.        
\end{enumerate}

\paragraph{Alternative multivariate distribution.} One could instead use the \citet{AzzaliniCapitanio2003} multivariate skew-$t$ distribution. However, its marginals are not univariate skew-$t$ with the specified parameters $(\xi_k, \omega_k, \alpha_k, \nu)$---the effective marginal skewness depends on the full parameter vector and correlation structure. Since we require exact control over each marginal to match the quantiles of the Reference and Baseline distributions, the copula approach is necessary.

\paragraph{Ignoring dependence.} Another alternative is to ignore dependence altogether and work only with the marginals. In this case, the IS weights for scenario $j=0,\ldots, J$ are obtained as $w_j^i \propto \prod_{k=1}^2\frac{p_{j,k}(\y^i)}{p_k(\y^i)}$. Potentially, one could even place different weights on the importance of matching the reference across variables by estimating the IS weights as $w_j^i \propto \prod_{k=1}^2\left(\frac{p_{j,k}(\y^i)}{p_k(\y^i)}\right)^{\gamma_k}$, where $\gamma_k \geq 0$ governs the importance assigned to matching that variable---setting $\gamma_k = 0$ excludes variable $k$ from the reweighting; $\gamma_k = 1$ assigns it full weight.

\subsubsection{Skew-$t$ copula sampling}\label{sec::skew_t_copula_sampling}
Without loss of generality, let $\mbf y = (y_1, y_2)'$ be a $2\times 1$ vector.
\begin{itemize}[label=$\square$,itemsep=0pt,parsep=0pt,topsep=0pt,leftmargin=1.4em]
    \item Each marginal $y_k$ follows a univariate skew-$t$ distribution: $y_k \sim \tS\tT(\xi_k, \omega_k, \alpha_k, \nu_k)$, where $\xi_k$ is location, $\omega_k$ is scale, $\alpha_k$ is skewness, and $\nu_k$ is degrees of freedom.
    \item The joint distribution is constructed using a copula with correlation matrix $\bm\rho$.
    \item To generate samples:
    \begin{itemize}[label=$\circ$,itemsep=0pt,parsep=0pt,topsep=0pt,leftmargin=1.25em]
        \item Generate $\mbf z \sim \tN(\mathbf{0}, \bm\rho)$ from multivariate normal with correlation $\bm\rho$.
        \item Transform to uniform: $u_k = \Phi(z_k)$ for $k=1,2$, where $\Phi$ is the standard normal CDF.
        \item Transform to skew-$t$ marginals: $y_k = F_k^{-1}(u_k)$, where $F_k$ is the skew-$t$ CDF with parameters $(\xi_k, \omega_k, \alpha_k, \nu_k)$.
    \end{itemize}
    \item This procedure guarantees exact marginal distributions $y_k \sim \tS\tT(\xi_k, \omega_k, \alpha_k, \nu_k)$ with specified correlation structure.
\end{itemize}

\paragraph{Copula choice.} In the empirical application we used a Gaussian copula for computational efficiency and because in our specific application the correlation structure is the same across all distributions. The $t$-copula could alternatively be used if tail dependence varies across distributions.

\subsubsection{Copula density evaluation}\label{sec:copula_density_evaluation}
For importance sampling, we require exact density evaluation $p(\mbf y)$ at sample points.
\begin{itemize}[label=$\square$,itemsep=0pt,parsep=0pt,topsep=0pt,leftmargin=1.4em]
    \item The joint density decomposes as:
    \[
       p(\mbf y) = c(F_1(y_1), F_2(y_2)) \times f_1(y_1) \times f_2(y_2),
    \]
    where $f_k$ are the skew-$t$ marginal densities, $F_k$ are the skew-$t$ marginal CDFs, and $c$ is the copula density.
    \item For Gaussian copula with correlation $\bm\rho$:
    \[
       c(u_1, u_2) = \frac{1}{\sqrt{|\bm\rho|}} \exp\Big\{-\frac{1}{2}\mbf z'(\bm\rho^{-1} - \mbf I)\mbf z\Big\},
       \quad \text{where } z_k = \Phi^{-1}(u_k).
    \]
    \item In log-space (for numerical stability):
    \[
       \log p(\mbf y) = \log c(F_1(y_1), F_2(y_2)) + \sum_{k=1}^2 \log f_k(y_k).
    \]
    \item This exact parametric density is used to compute importance weights $w^i = p(\y^i)/q(\y^i)$ for any two distributions $p$ and $q$ with known parameters.
\end{itemize}

\subsection{Calibrating the copula correlation matrix}
Let $\mbf y_{t+h}^{(m)} = (y_{1,t+h}^{(m)},\ldots,y_{K,t+h}^{(m)})'$, $m=1,\ldots,M$, denote the $m$th draw from the Large BVAR predictive distribution at horizon $h$ for the $K$ variables of interest.
\begin{itemize}[label=$\square$,itemsep=0pt,parsep=0pt,topsep=0pt,leftmargin=1.4em]    
    \item Stack the simulated vectors row-wise into the $M\times K$ matrix
    \[
       \Y_{t+h} =
       \begin{pmatrix}
       (\mbf y_{t+h}^{(1)})' &
       \ldots&
       (\mbf y_{t+h}^{(M)})'
       \end{pmatrix}'.
    \]
    \item The copula dependence matrix is defined as the sample Pearson correlation matrix across simulation draws:
    \[
       \bm\rho = \mathrm{corr}(\Y_{t+h}).
    \]
    \item This matrix is treated as fixed and is used in the Gaussian copula for the Reference, the Baseline, and all scenario distributions.
    \item Hence, the Large BVAR is used only to discipline the cross-sectional dependence structure, while the marginals are imposed separately using the fitted univariate skew-$t$ distributions.
\end{itemize}

%
%

\section{Predictive distribution for Core PCE price inflation}\label{sec:CPItoPCEPx}

OaR provides distributions for GDP growth and CPI inflation. Because the Tealbook scenarios are defined for core PCE price inflation, we convert the OaR CPI density into a core PCE density. We first shift the OaR CPI median by the difference between the Blue Chip core-PCE and CPI forecasts. We then rescale the CPI quantiles so that, for each symmetric quantile pair, the core-PCE-to-CPI interquantile-range ratio matches the corresponding ratio implied by the VAR forecast. This transformation preserves the asymmetry of the OaR CPI density in the constructed core-PCE reference.\smallskip

\leftskip 1em
\parindent -1em

\textsc{\color{Rosso} STEP 1: Scaling the median.} Let $\mathcal{Q}_{1,\tau}^R$ be the $\tau$-quantile from the Reference for core PCE price inflation, and $\mathcal{Q}_{2,\tau}^R$ be the same object for CPI inflation. Let $y_{1t+h}^{BC}$ be the Blue Chip forecast for core PCE price inflation. Then,
\[\mathcal{Q}_{1,50}^R=\mathcal{Q}_{2,50}^R+(y_{1t+h}^{BC}-y_{2t+h}^{BC})\]

\textsc{\color{Rosso} STEP 2: Scaling the other quantiles.}  Let $\mathcal{Q}_{1,\tau}^V$ be $\tau$-quantile from the empirical distribution of $y_{1t+h}^{(d)}$, where $y_{1t+h}^{(d)}$, $d=1,\ldots,D$ is the $d$-th draw of the unconditional forecast of core PCE price inflation from the VAR. Then, take the case of the $25$-th and $75$-th quantile. Let
\[\theta_1=(\mathcal{Q}_{2,\tau}^R-\mathcal{Q}_{2,(1-\tau)}^R)\frac{\mathcal{Q}_{1,\tau}^V-\mathcal{Q}_{1,(1-\tau)}^V}{\mathcal{Q}_{2,\tau}^V-\mathcal{Q}_{2,(1-\tau)}^V}\]
be the desired inter-quantile range in the core PCE price inflation Reference, and let
\[\theta_2=\frac{\mathcal{Q}_{2,\tau}^R+\mathcal{Q}_{2,(1-\tau)}^R-2\mathcal{Q}_{2,50}^R}{\mathcal{Q}_{2,\tau}^R - \mathcal{Q}_{2,(1-\tau)}^R}\]
be the $\tau$-quantile based skewness in the Reference for CPI inflation, then 
\begin{align*}
    \mathcal{Q}_{1,(1-\tau)}^R&=\mathcal{Q}_{1,50}^R+\frac{\theta_1(\theta_2-1)}{2}\\
    \mathcal{Q}_{1,\tau}^R&=\mathcal{Q}_{1,(1-\tau)}^R+\theta_1.      
\end{align*}
\leftskip 0em
\parindent 2em

%
%
\section{Extracting probabilities from the Blackbook}\label{sec:OpticalCharacterRecognition}

The probability estimates for the December 7, 2007 FRBNY Blackbook are approximate values obtain through Optical Character Recognition via ClaudeAI from the bar chart visualizations. To refine the first step estimate we provide Claude with the true SPF (Survey of Professional Forecasters) values. Specifically, we used the following three prompts to extract the numbers

\leftskip 1em
\parindent -1em

\textsc{\color{Rosso} Prompt 1:} Can you extract the numbers in the bars in the attached charts?

\textsc{\color{Rosso} Prompt 2:} Now I can provide you with the correct numbers for SPF so that you can revise those for FRBNY.

\textsc{\color{Rosso} Prompt 3:} Can you extract those numbers with more precision? say 1 decimal?

\leftskip 0em
\parindent 2em

Table \ref{tab:optical_character_recognition_results} shows the results we obtained from the different prompts.

\begin{table}\caption{Optical Character Recognition results}\label{tab:optical_character_recognition_results} \centering \small
\begin{tabular}{L{.2\textwidth}|C{.1\textwidth}C{.1\textwidth}|C{.1\textwidth}C{.1\textwidth}C{.1\textwidth}}\hline \hline
 & \multicolumn{2}{c}{SPF} \vline & \multicolumn{3}{c}{FRBNY} \\ \hline
Range &  1$^{st}$ e. &  True  &  1st e.   &  2nd e.   &  3rd e.   \\ \hline
$<$-2.0        &  ~0  &  ~0.22      &  ~4  &  ~4  &  ~4.5   \\
-2.0 to -1.0   &  ~0  &  ~0.44      &  ~3  &  ~3  &  ~3.5   \\
-1.0 to ~0.0   &  ~2  &  ~2.13      &  ~8  &  ~8  &  ~8.5   \\
~0.0 to ~1.0   &  ~7  &  ~7.28      &  18  &  18  &  18.0 \\
~1.0 to ~2.0   &  25  &  25.00      &  24  &  24  &  24.0 \\
~2.0 to ~3.0   &  45  &  45.02      &  21  &  20  &  20.5  \\
~3.0 to ~4.0   &  18  &  17.18      &  15  &  15  &  15.0 \\
~4.0 to ~5.0   &  ~3  &  ~2.10      &  ~6  &  ~6  &  ~5.5   \\
~5.0 to ~6.0   &  ~0  &  ~0.47      &  ~1  &  ~1  &  ~0.5   \\
$>$ 6.0        &  ~0  &  ~0.16      &  ~0  &  ~1  &  ~0.0  \\ \hline
\end{tabular}

\begin{tabular}{p{\textwidth}}\scriptsize
\textsc{Note:} The table compares probability values extracted from the Blackbook histogram using successive OCR prompts. For the SPF histogram, the column labeled ``True'' reports the probabilities implied by the published SPF distribution and is used as a benchmark. For the FRBNY histogram, ``1st e.'', ``2nd e.'', and ``3rd e.'' report the first, second, and third OCR-based extractions, respectively. All entries are percentages.
\end{tabular}
\end{table}

%
%
\section{How to convert annual-average GDP growth into quantile forecasts for Q4/Q4 GDP growth}\label{sec:annual_2_q4q4}

Let $i$ denote GDP growth and $r \in \{\text{BB}, \text{SPF}\}$ denote the two reference distributions. Each reference provides the set of $J$ quantile pairs shown in Table \ref{tab:optical_character_recognition_results}; we denote them as $\mathcal{Q}^{(i,r)} \;=\; \bigl\{(q_j^{(i)},\, p_j^{(i,r)})\bigr\}_{j=1}^{J},$ where $q_j^{(i)}$ is an outcome value (e.g., GDP growth of $2\%$) and $p_j^{(i,r)} = \Pr(\tilde{y}^{(i)} \leq q_j^{(i)})$ is the corresponding cumulative probability extracted from the Blackbook fan chart or the SPF histogram. Furthermore, let $Y_t$ be the $n$-dimensional vector of macroeconomic variables used to estimate the VAR.  The model is estimated with Bayesian methods with $N = 20{,}000$ posterior draws. From the VAR, we generate conditional forecasts $\{y_{T+h}^{(i,d)}\}_{d=1}^N$. Then, we converted annual-average GDP growth into quantile forecasts for Q4/Q4 GDP growth as follows:

\paragraph{STEP 1: Skew-$t$ fit the reference quantiles.} For each reference $r$, we fit a skew-$t$ distribution with parameters $\theta^{(r)} = (\xi, \omega, \alpha, \nu)$ on $\mathcal{Q}^{(i,r)}$ as in \citet{Adrianetal2019}. This delivers a smooth fitted \cdf $\hat{F}^{(r)}_{\tS\tT}$ and the corresponding \pdf $\hat{f}^{(r)}_{\tS\tT}$.

\paragraph{STEP 2: Calendar-year year-over-year growth rates} We construct annual-average growth rates from the quarterly VAR forecasts. Let $Y_t^{(i)}$ denote the log-level of variable $i$ in quarter $t$. The annual-average growth rate at quarter $t$ is defined as:
\begin{equation}
    \tilde{y}_t^{(i)} \;=\; \frac{1}{4}\sum_{k=0}^{3} Y_{t-k}^{(i)} \;-\; \frac{1}{4}\sum_{k=4}^{7} Y_{t-k}^{(i)}.
\end{equation}
Applied to the VAR forecasts, this delivers a $N$-draw sample $\{\tilde{y}_H^{(i,d)}\}_{d=1}^N$ of the CY-YoY growth rate at horizon $H = T+4$.

\paragraph{STEP 3: Optimal transport (quantile mapping)}

We tilt the $N$ VAR draws of the annual-average growth rate $\{\tilde{y}_H^{(i,d)}\}_{d=1}^N$ to match the reference marginal distribution using a rank-preserving (monotone) map. Let $\hat{F}^{(i)}_{\text{VAR}}$ denote the empirical \cdf of the VAR draws. For each draw $d$:
\begin{equation}
    \tilde{x}_{\text{ref}}^{(i,d)} \;=\; \bigl(\hat{F}^{(i,r)}_{\tS\tT}\bigr)^{-1}\!\!\left(\hat{F}^{(i)}_{\text{VAR}}\!\left(\tilde{y}_H^{(i,d)}\right)\right).
\end{equation}
This replaces the VAR marginal distribution with the reference distribution while preserving the rank ordering of draws, so that the joint dependence structure of the VAR is retained across variables.

\paragraph{STEP 4: Conditional BVAR forecasts:} The tilted annual draws $\{\tilde{x}_{\text{ref}}^{(i,d)}\}_{d=1}^N$ are imposed as hard constraints on the augmented VAR state space. The VAR is iterated forward draw-by-draw to recover the corresponding 4-quarter percent change:
\begin{equation}
    \Delta_4 y_H^{(i,d)} \;=\; y_H^{(i,d)} \;-\; y_{H-4}^{(i)}.
\end{equation}
This yields a draw sample $\{\Delta_4 y_H^{(i,d)}\}_{d=1}^N$ that is internally consistent with the VAR dynamics and anchored to the reference marginal at the annual frequency.

\paragraph{STEP 5: Quantiles estimation} A small set of quantiles is extracted from the $N$ draws at fixed probability levels $\{p_1,\ldots,p_K\}$:
$\hat{q}_k^{(i)} \;=\; \text{Quantile}\!\left(\bigl\{\Delta_4 y_H^{(i,d)}\bigr\}_{d=1}^N,\; p_k\right), \qquad k = 1,\ldots,K.$

\end{document}